\documentclass[a4paper,11pt]{article}
\usepackage{jheppub} % for details on the use of the package, please see the JINST-author-manual
\usepackage{lineno}
\usepackage{xcolor}
\usepackage{bm}
\usepackage{wrapfig,lipsum,booktabs}

\newcommand{\fig}[1]{{Fig.~\ref{#1}}}

\title{\boldmath Gravitational Duals from Equations of State}

\author[a,b]{Yago Bea,}
\author[b,c]{Raul Jimenez,}
\author[a,b,c]{David Mateos,}
\author[d]{Shuheng Liu,}
\author[d]{Pavlos Protopapas,}
\author[a,b]{Pedro Tarancón-Álvarez,}
\author[a,b]{Pablo Tejerina-Pérez}

\affiliation[a]{Departament de F\'\i sica Qu\'antica i Astrof\'\i sica,  Universitat de Barcelona, Mart\'\i\  i Franqu\`es 1, 
\mbox{ES-08028}, Barcelona, Spain.}
\affiliation[b]{Institut de Ci\`encies del Cosmos (ICC),  Universitat de Barcelona, Mart\'\i\  i Franqu\`es 1, 
\mbox{ES-08028}, Barcelona, Spain.}
\affiliation[c]{Instituci\'o Catalana de Recerca i Estudis Avan\c cats (ICREA), Passeig Llu\'\i s Companys 23, \\
\mbox{ES-08010}, Barcelona, Spain.}
\affiliation[d]{Institute for Applied Computational Science, Harvard University, Cambridge, MA.}

\emailAdd{yago.bea.b@gmail.com; raul.jimenez@icc.ub.edu; dmateos@fqa.ub.edu; pprotopapas@g.harvard.edu; shuheng\_liu@g.harvard.edu; pedro.tarancon@fqa.ub.edu;  pablo.tejerina@icc.ub.edu}

\abstract{Holography relates gravitational theories in five dimensions to four-dimensional quantum field   theories in flat space. Under this map, the equation of state of the field theory is encoded in the black hole solutions of the gravitational theory. Solving the five-dimensional Einstein's equations to determine the equation of state is an algorithmic, direct problem. Determining the gravitational theory that gives rise to a prescribed  equation of state is a much more challenging, inverse problem.  We present a novel approach to solve this problem based on physics-informed neural networks. The resulting algorithm is not only data-driven but also informed by the physics of the   Einstein's equations. We successfully apply it to theories with crossovers, first- and second-order phase transitions.}

\begin{document}
\maketitle
\flushbottom

\section{Introduction}
\label{sec:intro}
Holography \cite{Maldacena:1997re,Witten:1998qj,Gubser:1998bc} provides a valuable tool to study strongly coupled gauge theories with  a large number of colors. In this limit, the duality maps the quantum properties of the gauge theory to the classical properties of a gravitational theory in one dimension higher. This geometrization has profound consequences. A celebrated one is the fact that the equilibrium, thermodynamic properties of the gauge theory are encoded in properties of black hole horizons in the dual gravitational solutions. Not surprisingly, near-equilibrium aspects of the gauge theory, such as transport coefficients, are then encoded in small perturbations of the corresponding black hole horizons. Yet, the most dramatic consequence of the holographic map is that even the arbitrarily-far-from-equilibrium dynamics of the gauge theory can be determined by evolving in time the dual Einstein's equations. 

The equilibrium regime can be studied from first principles with conventional methods such as the lattice formulation of gauge theories. In contrast, in the far-from-equilibrium regime holography is often the only tool with which systematic, first-principle calculations are possible. A fruitful strategy is therefore  to construct a holographic model that reproduces the equilibrium properties of a gauge theory of interest, and then use the holographic side to study the far-from-equilibrium dynamics. From this perspective, one can think of holography as model building of a unique type, namely, one with the right to access the far-from-equilibrium regime. 

As we will see, implementing this strategy requires solving a challenging inverse problem. Early (semi)analytical attempts include \cite{Gubser:2008ny,Gubser:2008yx,Alanen:2009xs} (see \cite{Rougemont:2023gfz} for a review).  In this paper we will solve the problem using Physics-Informed Neural Networks (PINNs). With the exception of \cite{Hashimoto:2021ihd,Hashimoto:2022eij,Chen:2024ckb}, previous applications of NNs to holography \cite{Gan:2017nyt,You:2017guh,Hashimoto:2018ftp,Hashimoto:2018bnb,Hu:2019nea,Hashimoto:2019bih,Han:2019wue,Tan:2019czc,Yan:2020wcd,Akutagawa:2020yeo,Hashimoto:2020jug,Jokela:2020auu,Song:2020agw,Hashimoto:2021umd,Li:2022zjc,Jokela:2022fvh,Jokela:2023yun,Xu:2023eof,Park:2023slm,Ahn:2024gjf,Gu:2024lrz} have focused on the reconstruction of a specific \emph{solution} of the dual gravitational theory. In contrast, our algorithm reconstructs the gravitational \emph{theory} itself. 
Although  we will illustrate the method in a simple setup, we expect it to be generally applicable to inverse problems involving highly non-linear partial differential  equations. We envisage many potential applications 
that will be explored elsewhere. 

To illustrate our method, we will consider a four-dimensional Conformal Field Theory (CFT) deformed by a relevant operator $\mathcal{O}$. For concreteness, in this paper we will assume that $\mathcal{O}$ has conformal dimension three, but this is easily generalized. 
The holographic dual consists of five-dimensional gravity coupled to a scalar field $\phi$ with an appropriate potential $V(\phi)$ \cite{Erlich:2005qh,DaRold:2005mxj}. The existence of an ultraviolet (UV) fixed point in the gauge theory translates into  anti-de Sitter space (AdS) boundary conditions on the gravity side. Despite their simplicity, this type of models capture the essence of many fundamental gauge theory properties such as thermal phase transitions \cite{Witten:1998zw}, confinement \cite{Karch:2006pv}, etc. 

All the properties of the gauge theory, in particular its thermodynamics,  are encoded in the scalar potential $V(\phi)$.  Given the potential, in order to determine the thermodynamics one needs to solve a \emph{direct} problem. This consists of finding all the static, regular black hole solutions of the Einstein--Klein--Gordon (EKG) equations. This set of solutions can be conveniently parameterized by the value of $\phi$ at the horizon, $\phi_H$. For each solution one  computes the area density and the surface gravity of the horizon. From them one  extracts the Bekenstein-Hawking entropy density $S(\phi_H)$ and the Hawking temperature $T(\phi_H)$. These quantities are then identified with the entropy density and the temperature of the corresponding thermal state in the gauge theory. The result is an equation of state $S(T)$. Any other thermodynamic property such as the energy density, the pressure, etc. can be obtained from $S(T)$ via thermodynamic identities. 

Since the scalar potential enters the EKG equations that must be solved to determine the thermodynamics, it is clear that the function $S(T)$ is a  functional of the function $V(\phi)$. The determination of $S(T)$ from  $V(\phi)$ is a direct problem because it can be done algorithmically, as we have just described. The \emph{inverse} problem consists of finding a potential $V(\phi)$ that reproduces a given  $S(T)$. The fact that the EKG equations are highly non-linear differential equations makes this an extremely challenging problem.  We will solve it using PINNs. These NNs are uniquely suited for  inverse problems for two reasons. First, they use the differential equations when evaluating the loss of the NN. Second, they allow for the reconstruction of a general function by encoding it as a second NN. As we will see, the role of the equation of state is to provide the necessary physical input to solve these equations, namely the boundary conditions for the EKG equations.
We will illustrate the applicability of the method by reconstructing gravitational theories that give rise to equations of state with crossovers, first- and second-order phase transitions.

\section{Holography and the direct problem}
\label{holomodel}
As explained above, we will focus on an EKG theory in five dimensions. The action takes the form
\begin{equation}
    S=\frac{2}{\kappa_5^2} \int dx^5\sqrt{-g}\left[\frac{1}{4} R -\frac{1}{2}(\nabla \phi)^2 - V(\phi) \right] \,,
    \label{EinsteinScalarAction}
\end{equation}
where $\kappa_5$ is the gravitational coupling
and $V(\phi)$ is the scalar potential. 

The thermodynamics of the gauge theory is determined by finding all the static, planar black hole solutions of  \eqref{EinsteinScalarAction}. Because of their translation invariance along the gauge theory directions, these solutions are usually referred to as ``black brane'' solutions. In order to find them, we consider an asymptotically AdS spacetime and we choose coordinates $t, x, y, z$ for the Minkowski metric at the boundary. We then write the bulk metric in Eddington-Finkelstein coordinates as
\begin{equation}
ds^2=-A dt^2+\Sigma^2(dx^2+dy^2+dz^2)-\frac{2}{u^2}\, dt du \,\, ,
\label{metric_ansatz}
\end{equation}
where $u$ is the null holographic coordinate. The AdS boundary is located at $u=0$. This region corresponds to the UV of the gauge theory. The metric functions $A$ and $\Sigma$, as well as the scalar field $\phi$, are functions of $u$. Substituting the ansatz \eqref{metric_ansatz} into the equations of motion derived from \eqref{EinsteinScalarAction} we obtain four ordinary differential equations (ODEs)
\begin{subequations}
 \label{Einstein-scalar_equations}
\begin{eqnarray}
 \Sigma^{\prime \prime}+\frac{2}{u} \Sigma^{\prime}+\frac{2}{3} \Sigma \phi^{\prime 2}  &= & 0  \,\, , \\[2mm]
A^{\prime \prime}+\frac{8}{3 u^4} V(\phi)+A^{\prime}\left(\frac{2}{u}+\frac{3 \Sigma^{\prime}}{\Sigma}\right)  &= & 0  \,\, , \\[2mm]
\phi^{\prime \prime}-\frac{1}{u^4 A} \frac{\partial V(\phi)}{\partial \phi}+\phi^{\prime}\left(\frac{2}{u}+\frac{A'}{A}+3 \frac{\Sigma^{\prime}}{\Sigma}\right) & = &0  \,\, , \\[2mm]
 A^{\prime}+2 A \frac{\Sigma^{\prime}}{\Sigma}-\frac{2 \Sigma}{3 \Sigma^{\prime}}\left(A \phi^{\prime 2}-\frac{2}{u^4} V(\phi)\right) & = & 0 \,\, ,
 \label{constraint}
\end{eqnarray}
\end{subequations}
where a prime indicates differentiation  with respect to $u$. The last equation is first-order in derivatives because it is a constraint associated to our gauge choice for the $u$-coordinate. As a consequence, the four equations \eqref{Einstein-scalar_equations} are not all independent: two of the second-order equations plus the constraint imply the third second-order equation. This redundancy will be useful for our PINN. 

The equations of motion must be solved subject to appropriate boundary conditions. Near $u=0$, these ensure that the geometry is asymptotically AdS and take the form 
\begin{subequations}
\label{UVexpansion}
\begin{eqnarray}
 A(u)  &= & \frac{1}{u^2}+ \cdots   \,\, , \label{UVexpansion_a}\\[2mm]
\Sigma(u) &= & \frac{1}{u}+ \cdots  \,\, , \label{UVexpansion_b}\\[2mm]
\phi(u) & = & \Lambda u + \cdots  \,\, , \label{UVexpansion_c} 
\end{eqnarray}
\end{subequations} 
where the dots indicate subleading terms in the limit $u\to 0$. In the expressions for the metric functions we have fixed some integration constants by setting to unity the coefficients of the leading terms. As anticipated above, in the boundary condition for the scalar field we have assumed that the dual scalar operator, $\mathcal{O}$, has conformal dimension 
\begin{equation}
    \Delta=3 \,.
    \label{delta}
\end{equation} 
This means that the constant $\Lambda$ has mass dimension 1.

For a static black brane, the horizon is characterized by a simple zero in the $g_{tt}$ component of the metric, namely in  $A(u)$, at some position $u=u_H$. Performing a Taylor expansion around this point we obtain 
\begin{subequations}
\label{Conditions_horizon}
\begin{eqnarray}
 A(u)  &= & A_1 (u_H-u)+ \cdots  \,\, , 
 \label{Conditions_horizon_a}\\[2mm]
\Sigma(u) &= & \Sigma_0 + \Sigma_1 (u_H-u) + \cdots  \,\, , \label{Conditions_horizon_b}\\[2mm]
\phi(u) & = & \phi_H + \phi_1 (u_H-u)+ \cdots  \,\,. \label{Conditions_horizon_c} 
\end{eqnarray}
\end{subequations}
Reparametrization  invariance allows us to set the horizon of the black brane at $u_H=1$, which we will do hereafter. 
In these equations we have  set to zero some integration constants in order to ensure  regularity at the horizon. This choice, together with the choice of integration constants in \eqref{UVexpansion}, implies that all the coefficients in \eqref{Conditions_horizon} except for $\phi_H$ are fixed by the requirement that a solution of the form \eqref{Conditions_horizon} at the horizon  matches a solution of the form \eqref{UVexpansion} at the boundary. It follows that there is a one-to-one map between the value of the scalar field at the horizon   and physically different black brane solutions. This is consistent with the expectation that the different equilibrium states of the gauge theory are characterized by a single parameter, namely the temperature $T$. This and the entropy density can be computed from the surface gravity and the area density of the horizon as 
\begin{subequations}
\label{temperature_entropy}
\begin{eqnarray}
 \frac{T}{\Lambda}  &= & \frac{1}{4 \pi}\frac{A_1}{\Lambda}    \,\, , \label{temperature_entropy_a}\\[2mm]
\frac{S}{\Lambda^3} &= & \frac{2 \pi}{\kappa^2} \left(\frac{\Sigma_0 }{\Lambda}\right)^3  \,\, . 
\label{temperature_entropy_b}
\end{eqnarray}
\end{subequations}
In these equations we have assumed \eqref{delta} again.  The scale $\Lambda$ is the only intrinsic scale in the gauge theory, so we will work in units in which 
\begin{equation}
    \Lambda =1 \,.
    \label{lll}
\end{equation}
Similarly, on the gravity side it will be convenient to work in units such that $\kappa^2=2$.  

We can now state  what  solving the direct problem means. The only input is some scalar potential $V(\phi)$. Given this function, we must solve the EKG equations \eqref{Einstein-scalar_equations}. A standard way to do this is to numerically integrate the equations from the horizon to the boundary. For each value of $\phi_H$ there is a unique choice of $A_1$ and $\Sigma_0$ such that the solution near the boundary takes the form \eqref{UVexpansion}. This requirement determines $A_1$ and $\Sigma_0$ in terms of $\phi_H$. 
In Figs.~\ref{fig:solutionsASigmaPhiM1} and \ref{fig:solutionsASigmaPhiphiM5} we show the functions   $A(u)$, $\Sigma(u)$ and $\phi(u)$ for three different black brane solutions for two different gauge theories. 
Substituting the values of $A_1(\phi_H)$ and $\Sigma_0(\phi_H)$ in \eqref{temperature_entropy} we  obtain $T(\phi_H)$ and $S(\phi_H)$ and, as a consequence, the thermodynamic curve $S(T)$. Note that, while there is always a single state for each value of $\phi_H$, in theories with phase transitions there can be more than one state for a given value of $T$. In other words, while the map from $\phi_H$ to equilibrium states is single-valued, the map from $T$ can be multi-valued.

\section{Inverse problem}
\label{inv}
It is clear from Sec.~\ref{holomodel} that solving the direct problem is algorithmic. Solving the inverse problem is not. In the inverse problem the only input is the thermodynamic curve $S(T)$. Through equations \eqref{temperature_entropy}, each pair of values $(T, S)$ determines a pair $(A_1, \Sigma_0)$. In turn, each of these pairs  provides boundary conditions at the horizon through equations \eqref{Conditions_horizon}. Given this knowledge, the inverse problem consists of finding a unique potential $V(\phi)$ such that the EKG equations with this same potential admit black brane solutions for all these boundary conditions.

\begin{figure}
    \centering
    \includegraphics[width = 0.93\columnwidth]{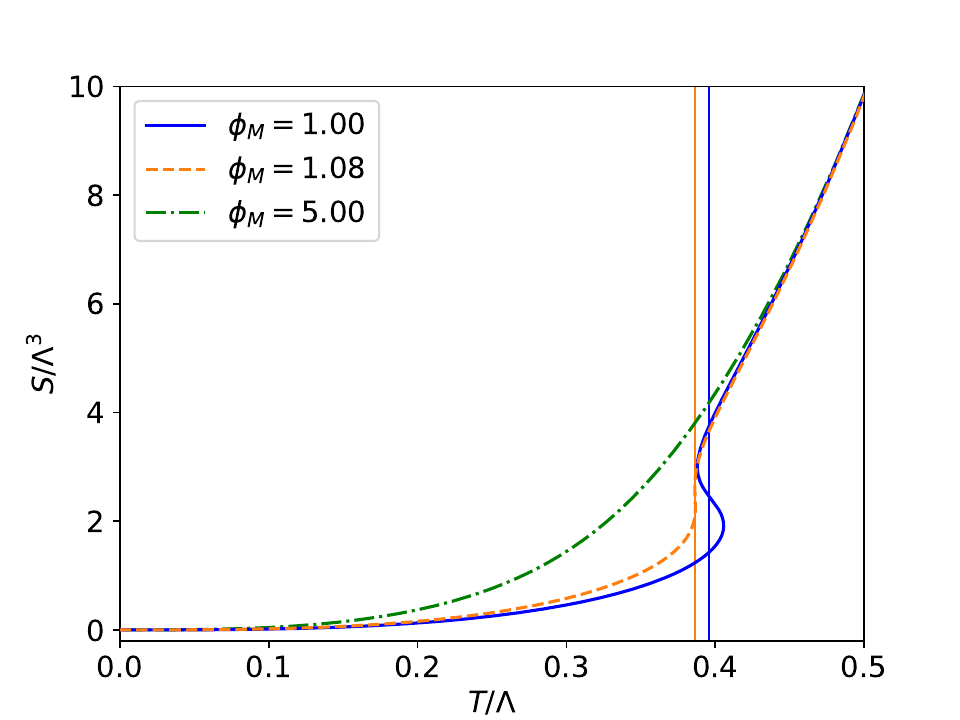}\\
    \includegraphics[width = 0.95 \columnwidth]{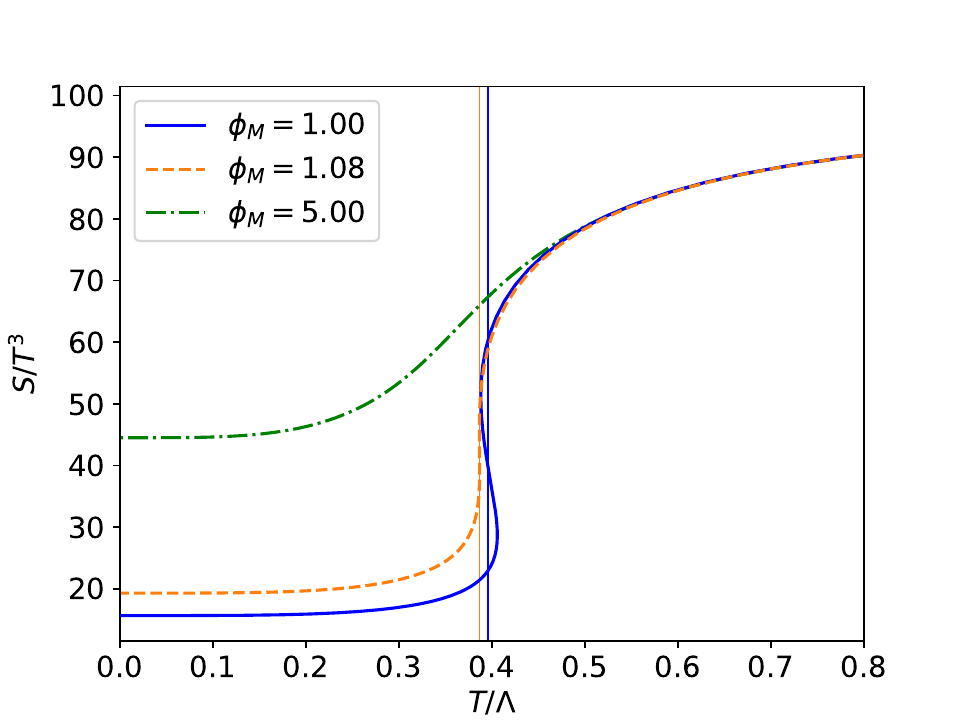}
    \caption{Equation of state for a gauge theory with a crossover ($\phi_M=5$), a second-order phase transition  ($\phi_M=1.08$), and a first-order phase transition ($\phi_M=1$) \cite{Bea:2018whf}. In the latter two cases, the critical temperatures are indicated by  solid vertical lines. The entropy density is shown in units of the intrinsic scale in the theory (top) and in units of the temperature (bottom).}
    \label{SofT}
\end{figure}

Our goal is thus to construct a PINN able to solve this inverse problem. In order to test its accuracy, we will ask the PINN to find the potentials for a family of equations of state $S_{\rm{theory}}(T)$ constructed in \cite{Bea:2018whf} and labeled by one parameter dubbed $\phi_M$. As an illustration, \fig{SofT} shows the equations of state for $\phi_M=5,1.08$ and 1. 
The reason why we choose these examples is that they cover the three possible cases of a crossover, a second-order or a first-order phase transition. 

In the high-temperature limit, the entropy density behaves as 
\begin{equation}
    S_{\rm{theory}}(T) = \pi^4 \, T^3 + 
    \frac{3 \pi^4}{64}\,  T +\cdots \,.
    \label{leadingStheory}
\end{equation}
Comparing with Eq.~\eqref{leadingS}, we see that this corresponds to a CFT deformed by an operator $\mathcal{O}$ of dimension \eqref{delta}, in agreement with the boundary condition \eqref{UVexpansion_c} that we imposed on the dual scalar field $\phi$. Solving the EKG equations \eqref{Einstein-scalar_equations} perturbatively near the boundary, it can be shown that, in our conventions, the form of the entropy density \eqref{leadingStheory} implies the following  conditions on the scalar potential at $\phi=0$:
\begin{equation}
    C_0\equiv V(0)+3 =0  \,,\qquad 
    C_1\equiv V'(0)=0 \,,\qquad 
    C_2\equiv V''(0)+3 =0 \,.
    \label{Vconditions}
\end{equation}
The fact that these conditions can be imposed at $\phi=0$ can always be achieved by a shift in $\phi$. The first condition fixes the coefficient of the $T^3$ term in $S_{\rm{theory}}$. The last two conditions state that $\phi=0$ is a maximum of the potential. On the gravity side, this corresponds to the AdS boundary condition on the metric, and on the gauge theory side it corresponds to the presence of an UV fixed point, which in turn fixes the leading $T^3$ scaling in $S_{\rm{theory}}$. Finally, the last condition imposes that the dimension of the gauge theory operator dual to $\phi$ is precisely \eqref{delta}, in agreement with the fact that the leading correction in the entropy density is linear in $T$. Since all this information is contained in the equation of state, which is assumed to be given, we will provide our PINN with these boundary conditions on the potential. All the results reported in Sec.~\ref{results} were obtained under these assumptions. Nevertheless, we have also performed tests in which we do not provide the PINN with the information in \eqref{Vconditions}. We will comment on these in Sec.~\ref{disc}.

By feeding the different $S_{\rm{theory}}(T)$ to the PINN we obtain a family of reconstructed potentials $V_{\rm{PINN}}(\phi)$. We emphasize that, other than the boundary conditions \eqref{UVexpansion}, 
\eqref{Conditions_horizon} and \eqref{Vconditions}, we do not supply the PINN with \emph{any} information about the functional form of the potential or of the metric functions. We then solve the direct problem with each of these $V_{\rm{PINN}}(\phi)$ and obtain a set of thermodynamic curves $S_{\rm{PINN}}(T)$. In Sec.~\ref{results} we will compare $S_{\rm{PINN}}(T)$ with $S_{\rm{theory}}(T)$. Needless to say, the success of the method requires that these two curves be as close as possible. 

\begin{figure}
    \centering
    \includegraphics[width=0.98\textwidth]{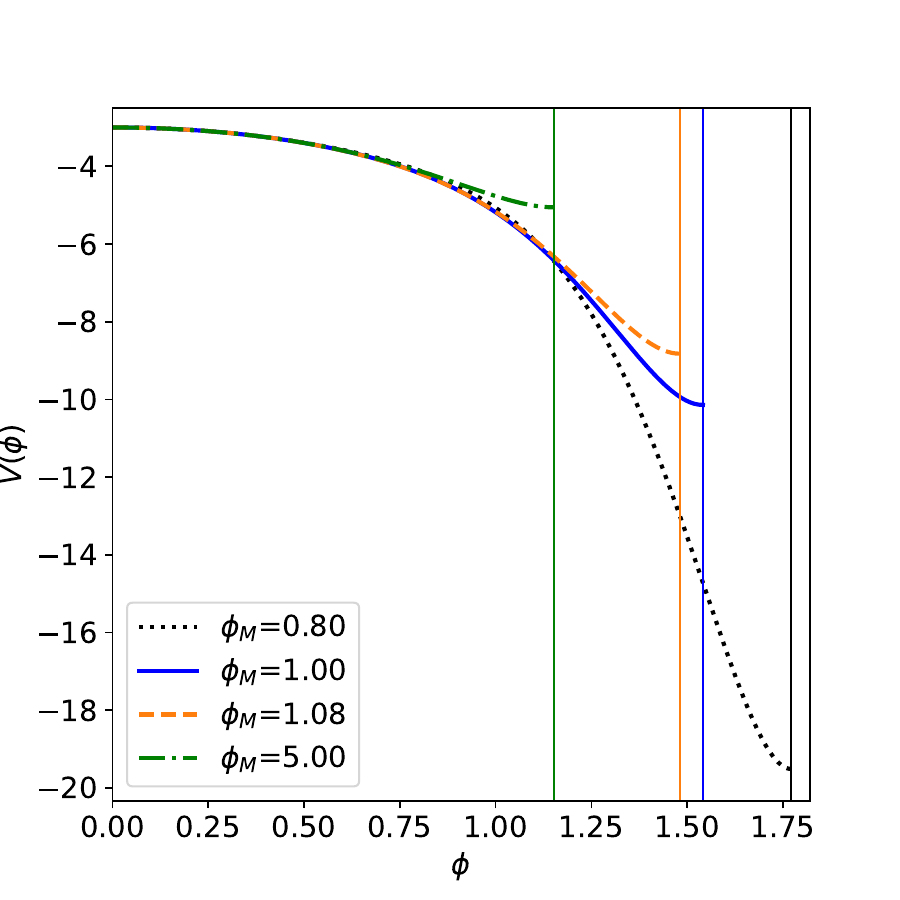}
    \caption{Potentials that give rise to the equations of state in \fig{SofT} \cite{Bea:2018whf}. All potentials share the same maximum at $\phi=0$, but they have minima at different positions indicated by the solid vertical lines.}
    \label{fig: V for different phiM}
\end{figure}

The family of equations of state $S_{\rm{theory}}(T)$ were constructed in \cite{Bea:2018whf} by solving the direct problem for the family of potentials given by
\begin{equation}
    V_{\rm{theory}}(\phi)= 
    -\frac{4}{3}\mathcal{W}(\phi)^2 + 
\frac{1}{2} \mathcal{W}'(\phi)^2 \,,
\label{Vtheory}
\end{equation}
where
\begin{equation}
    \mathcal{W}(\phi)=-\frac{3}{2} - \frac{\phi^2}{2}
    - \frac{\phi^4}{4\phi_M^2} + \frac{\phi^6}{10} \,.
\end{equation}
Some of these are plotted in \fig{fig: V for different phiM}. The maxima of these potentials corresponds to the UV fixed point in the gauge theory. These potentials also exhibit a minimum at some $\phi>0$, corresponding to the presence of an IR fixed point in the dual gauge theory. We will see that the PINN is able to discover this feature.  

Knowledge of the true potentials means that, in Sec.~\ref{results}, we will be able to perform an additional, intermediate test: we will be able to compare the potential reconstructed by the PINN with the true potential. This will illustrate a crucial property of this inverse problem, namely the fact that the potential must be reconstructed with very high precision in order for the method to be successful.

\section{Methodology}
\label{sec: methodology}
 In this section we describe our procedure to solve the EKG equations and reconstruct the potential non-parametrically, i.e.~as a free-form function, using our implementation of PINNs to solve ODEs. First, we review some general aspects about the use of NNs  to solve differential equations.

\subsection{Background}
A neural network (NN) is a machine-learning algorithm that produces a structure of interconnected nodes, or neurons,  disposed in a layered structure, which can be trained to solve a certain task through minimization of a loss function. In a feedforward NN, which is the type of  architecture that we consider in this work, a neuron takes an input and affine-transforms it, passes the resulting transformed data through an activation function, and produces an output. Each neuron in a layer is connected to all neurons in the following layer. Every connection carries a weight $w_i$ that multiplies the output of the neuron of the previous layer passing via some non-linear function to the $i$-th neuron in the next layer. These weights are the main parameters of the NN followed by the activation function. The final output of the NN will depend on the entire set of all weights, which we denote by $W_\text{net}$. A NN can be trained so that its output gives an approximate solution to a certain problem. How well the NN's output resembles the exact solution is encoded in the loss function. By changing the parameters $W_\text{net}$ in a way that minimizes this loss function,\footnote{That is, flowing towards (or against) the steepest gradient of the loss function represented in a high dimensional parameter space.} the NN will, in principle,  approximate increasingly well the solution to the problem it is being trained on. It is worth noting that, according to the universal approximation theorem, NNs with a single hidden layer containing a finite number of neurons can approximate any continuous function to arbitrary accuracy.  In the case of PINNs, the NN trains on the differential equations it is trying to solve. As a consequence, once trained, the PINN itself becomes the approximate solution.

Our implementation to solve the inverse problem is relatively new and original for this kind of problems, but it exploits previously known PINNs. These are deep neural networks (DNNs) that can approximate the solution of differential equations (DEs) and initial and boundary values. The early results of Dissanayake and Phan-Thien \cite{dissanayake_neural-network-based_1994}  and then by Lagaris \cite{lagaris_original_PINN}, solving DEs using NNs, were followed by a rapid growth of publications on PINNs in recent years. PINNs have many advantages compared to traditional numerical solvers. 
They provide closed-form solutions, eliminating the need for iterative solvers and reducing computational overhead. PINNs are mesh-free, allowing for on-demand solution computation post-training, which enhances efficiency in solving complex problems. Their ability to leverage transfer learning enables rapid discovery of new solutions by transferring knowledge from related tasks, enhancing adaptability and accelerating convergence. PINNs are invertible, which makes them suitable for inverse problems where obtaining the original input from the output is essential. Additionally, PINNs can be parametrized to include parameters of the differential equation as input.

Many dynamical systems are described by ODEs which relate the rates and values of state variables and external driving functions. While some simple ODEs have closed-form solutions, the vast majority must be solved approximately using discretization of the domain or via spectral methods \cite{bernardi_spectral_1997}. The former approximating methods are more general, with Runge-Kutta and multi-step methods as typical examples. These methods seek to numerically integrate the ODEs starting from a value of the independent variable at a boundary and stepping away. These conventional methods are generally efficient in determining the state of a system for a sequence of values. However, if we are only interested in the state at a specific later value, substantial computational effort must still be spent determining all the states at steps leading up to the state of interest. This ordering also limits parallelizability of the conventional single- and multi-step methods because, 
until the preceding state is known, processors tasked with finding a segment of the system's evolution over a later interval cannot start calculating the correct piece of the trajectory.

The idea of approximating DE solutions with NN was first developed by Dissanayake and Phan-Thien, where training was carried out by minimizing a loss based on the network's satisfaction with the boundary conditions and DEs themselves \cite{dissanayake_neural-network-based_1994}. Lagaris \textit{et al.}~\cite{lagaris_reparametrization} showed that the form of the network could be chosen to satisfy the boundary conditions by construction, and that automatic differentiation could be used to compute the derivatives appearing in the loss function \cite{lagaris_artificial_1998}. After training, the approximate value of the solution at any point within the training range can be computed without having to compute the previous states first. This method has been extended to systems with irregular boundaries \cite{lagaris_neural-network_2000,mcfall_artificial_2009}, applied to solving partial differential equations (PDE) arising in fluid mechanics \cite{baymani_artificial_2010} and Hamiltonian systems \cite{ShanMarios}, and software packages have been developed to facilitate its application \cite{lu_deepxde_2020,koryagin_pydens_2019,chen_neurodiffeq_2020}. In the Lagaris approach, the NN learns a single solution to the ODE. For different sets of initial conditions or for different sets of parameters in the differential equation, the network has to be re-trained on the new task.

One of us (PP) has developed an extension of the Lagaris method \cite{lagaris_artificial_1998} where the neural network is instead taught a variety of solutions to a parameterized differential equation. This increases the reusability of the trained network and can speed up tasks that require knowing many solutions, such as for Bayesian parameter inference, propagating uncertainty distributions in dynamical systems, or inverse problems. It is straightforward to extend this approach to problems containing various types of boundary conditions, to PDEs and higher derivatives. Here, we focus on boundary-condition problems in a system of first-order ODEs. We will show that our method has promise when applied to a variety of tasks requiring quick, parallel evaluation of multiple solutions to an ODE, and where it is useful to be able to differentiate the state at a particular value of the evolution variable with respect to the initial conditions or ODE parameters.

\subsection{Solution bundles}
\label{bundle}

When working with ODEs it is common to require multiple solutions corresponding to different initial and boundary values. In dynamical systems, each of these solutions represents a trajectory, tracing out an alternate time evolution of the state of the system. Similar situations apply when we have multiple boundary values. In addition, when an ODE depends on a parameter, for example, a physical constant whose value has an associated uncertainty, it can be useful to have different solutions for different values of the parameter.

In our case, we have a range of boundary conditions: the range of values of $S(T)$. Finding the free-form function $V(\phi)$ requires information from all points on the curve, that is, having information about {\em all} boundary conditions at the same time. Thus, bundle solving is an essential ingredient to solve our problem.

\subsection{Solving the Einstein-Klein-Gordon equations}

\subsubsection{ODE setup}
We now proceed to implement the EKG equations \eqref{Einstein-scalar_equations} in the NN. For this purpose, it is convenient to redefine the functions $A$, $\Sigma$, $\phi$ in such a way that the solutions are finite in the entire computational domain $u\in\left[0,1\right]$. In view of the asymptotic behavior \eqref{UVexpansion}, 
this can be achieved through the following redefinitions:
\begin{equation}
\tilde{\Sigma} = u \, \Sigma \,\, ,
 ~~~~   
\tilde{A} = u^2 A  \, .
\label{RedefinitionsASigma}
\end{equation}
It is also convenient to transform the EKG equations into first-order ODEs by thinking of the first-derivatives of $\tilde{A}, \tilde{\Sigma}$ and $\phi$ as independent variables. Together with the constraint \eqref{constraint}, this results in 6 functions ($\nu_{\Sigma}  , \nu_A, \nu_{\phi}, \tilde{\Sigma}, \tilde{A}, \phi$) subject to the following seven, coupled, first-order ODEs 
\begin{equation}
    E_\alpha=0 \,,\qquad \alpha=1, \ldots, 7\,,
    \label{Ealpha}
\end{equation}
where
\begin{subequations}
\label{1s_order_ODEs}
\begin{eqnarray}
E_1 & = & \nu_{\Sigma} - \tilde{\Sigma}'   \,,  \\[2mm]
E_2 & = & \nu_A - \tilde{A}' \,, \\[2mm]
E_3 & = & \nu_\phi - \phi'  \,, \\[2mm]
E_4 & = & \nu_{\Sigma}' +  \frac{2}{3} \tilde{\Sigma} \, 
 \nu_{\phi}^2  \,,\\[2mm]
E_5 & = & u^2 \, \tilde{\Sigma} \, \nu_A^{\prime} + \frac{8}{3} \, V(\phi)\, \tilde{\Sigma}+ 
    \nu_A \, \left(  3u^2 \, \nu_{\Sigma} - 5u\, \tilde{\Sigma}  \right) + 
           \tilde{A} \left( 8\tilde{\Sigma} - 6u \,\nu_{\Sigma} \right) \,, \\[2mm] 
E_6 & = & u^2 \,\tilde{\Sigma} \, \tilde{A}\, \nu_\phi' - \tilde{\Sigma}\,  \frac{d V}{d \phi} 
        + \nu_{\phi} \left( -3 u \, \tilde{A} \, \tilde{\Sigma}  + u^2 \,\tilde{\Sigma} \,\nu_A 
        + 3 u^2\, \nu_{\Sigma} \, \tilde{A} \right)  \,,  \\[2mm]     
E_7 & = & \left( u \, \nu_{\Sigma}-\tilde{\Sigma} \right) 
\left( u^2 \, \tilde{\Sigma} \, \nu_A + 2 u^2  \, \tilde{A} \, \nu_{\Sigma} - 4 u \tilde{A} \tilde{\Sigma} \right)
- \frac{2}{3}  u \, \tilde{\Sigma}^2 
\left( u^2 \tilde{A} \, \nu_\phi^2 - 2 V(\phi) \right) \,. \,\,\,\,\,\,       
\end{eqnarray}
\end{subequations}
In terms of these variables, the boundary conditions discussed in Sec.~\ref{holomodel} translate into 
\begin{subequations}
    \label{BoundaryConditions}
    \begin{align}
    \tilde{A}|_{u=0} &=1 \,\,,  \label{BoundaryCondition_a}\\[1mm]
\tilde{\Sigma}|_{u=0} &=1 \,\,, \label{BoundaryCondition_b} \\[1mm] 
\phi |_{u=0} &=0 \,\,,  \label{BoundaryCondition_c}\\[1mm]
\nu_\phi|_{u=0} &= 1 \,\,, \label{BoundaryCondition_d}\\[1mm]
\tilde{A}|_{u=1} &=0 \,\,,  \label{BoundaryCondition_e}\\[1mm]
\nu_A |_{u=1} &= - 4 \pi T \,\,, \label{BoundaryCondition_f}\\[1mm]
\tilde\Sigma_{u=1} &= \left( S/\pi \right)^{1/3}\,\,.
\label{BoundaryCondition_g}
    \end{align}
\end{subequations}
The conditions (\ref{BoundaryCondition_a})-(\ref{BoundaryCondition_c}) correspond to imposing that the solution approaches AdS at $u=0$, in agreement with \eqref{UVexpansion}. Condition \eqref{BoundaryCondition_d} corresponds to imposing the fact that the dual operator has dimension \eqref{delta} and the convention \eqref{lll} for the scale of the boundary theory.
The condition \eqref{BoundaryCondition_e} corresponds to imposing that the horizon is at $u=1$, consistently with \eqref{Conditions_horizon_a}. 
Finally, (\ref{BoundaryCondition_f}) and (\ref{BoundaryCondition_g}) correspond to conditions (\ref{Conditions_horizon_b}), (\ref{Conditions_horizon_c}) and \eqref{temperature_entropy}.

In the direct problem, we choose a value of the scalar field, we obtain the numerical solution of the black brane, and we read off the entropy and the temperature. In contrast, in the inverse problem we start from the equation of state $S(T)$ and use it as a boundary condition: $S$ and $T$ in (\ref{BoundaryCondition_f})-(\ref{BoundaryCondition_g}) are related by the equation of state. In the inverse problem we do not impose a condition for the scalar field at the horizon. The reason is that this information is not directly contained in the equation of state but can only be determined in combination with the potential itself.

\subsubsection{NN setup and procedure} \label{subsubsec: NN setup and procedure}
\begin{figure}[t]
    \centering
\includegraphics[width=\textwidth]{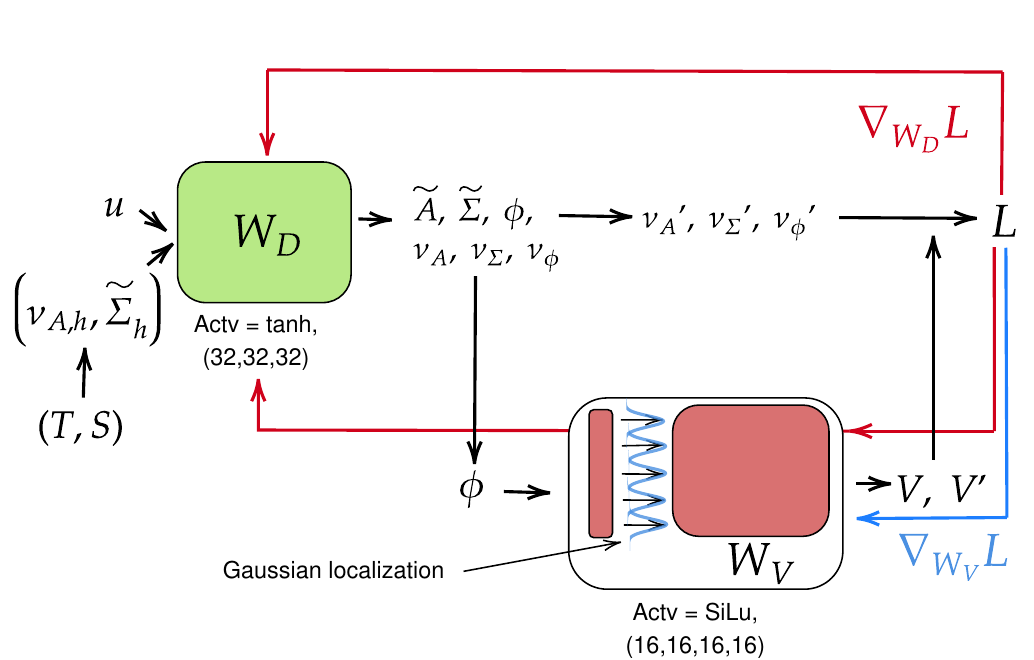}
    \caption{Structure of the implemented NN-system of two NNs with Gaussian localization. The red and blue arrows  represent the backpropagation process, in which the parameters $W_j$ ($j=D,V$) are modified such that the NN's state in parameter space flows in the direction of the gradient of the loss function, $\nabla_{W_j}L$.}
    \label{fig:gaussian architecture}
\end{figure}

We use a dual NN strategy to solve the problem --- see \fig{fig:gaussian architecture}. 
Although this architecture is computationally more demanding than having only a single NN, it is better suited for our problem for several reasons. The first and most important one is that we want the reconstructed potential to have information about the entire set of boundary conditions. Using two separate NNs, we allow NN-V to find the function that better fits all the given boundary conditions along the $S(T)$ curve. Another interesting feature of using a dual network architecture is that the response of the potential NN can be much more non-linear. This is an important property when the reconstructed function depends on the boundary conditions in a highly non-linear way. Our method is organized into focused steps, making it manageable and easier to interpret. 

\begin{enumerate}
    \item The neural network NN-Solver with parameters \( W_D \) takes as an input a discrete set of  pairs  \( (T_i, S_i) \), with $T_i$ and $S_i$ related by the equation of state,  along with a discrete set of values $u_n$ for the $u$-coordinate, with $u_n \in [0,1]$. This network then produces estimates for the dependent variables, namely for the functions $\tilde{A}$, $\tilde{\Sigma}$, ${\phi}$, $\nu_A$, $\nu_{\Sigma}$ and \(\nu_\phi \). By construction, these functions  obey the boundary conditions \eqref{BoundaryConditions} exactly. This is accomplished as described in \cite{chen_neurodiffeq_2020}.  Each point in the equation of state, namely each pair $(T_i, S_i)$,  corresponds to an entire geometry on the gravitational side described by the functions above. Thus, these functions should be considered to depend on one coordinate, $u$, and one parameter, the pair $(T,S)$. Therefore, we will write  expressions such as $\phi(u, (T,S))$, and similarly for the other functions. In some cases, specifying $T$ and $S$ separately is redundant, but not in others. Therefore, we will keep track of both $T$ and $S$. The derivatives with respect to the $u$-coordinate, \( \nu_A' \), \( \nu_{\Sigma}' \), and \( \nu_\phi' \), are computed analytically using automatic differentiation.
    
    \item The set of predicted values of the scalar field, $\phi_{i, n}=\phi(u_n, (T_i, S_i))$,  are then fed into a second neural network,  NN-V, with parameters $W_V$. This network predicts \( V(\phi_{i,n}) \) based on these values. Automatic differentiation is also used to find the derivative of \( V(\phi) \) with respect to \( \phi \).    

    \item These predicted values are  substituted back into the original set of ODEs \eqref{1s_order_ODEs}. We then compute and minimize a loss function $L$ over all the different weights. This minimization is achieved by fine-tuning the parameters \( W_D \) and \( W_V \) through stochastic gradient descent. The loss function is 
    defined as 
     \begin{equation}
     L\, \equiv \, \sum_\alpha \sum_{n} \sum_{i} \, E_\alpha \Big( u_n, (T_i, S_i) \Big)^2 
     + \lambda \Big( C_0^2 + C_1^2 +C_2^2 \Big)\,.
    \label{loss}
     \end{equation}
     The first term is the sum of the residues of all equations of motion \eqref{Ealpha} (sum over $\alpha$), evaluated on all the values of the $u$-coordinate (sum over $n$), and for all the different solutions (sum over $i$). The second term is the sum of the three boundary conditions \eqref{Vconditions} on the potential. The larger the coefficient of this term, the more efficiently these conditions are enforced by the NN. We choose an initial value of $\lambda= 0$ for the first million epochs. 
     After that, we set $\lambda = 50$ to enforce the boundary conditions \eqref{Vconditions}, and then train the model for an additional $(0.5-1)\cdot 10^{6}$  epochs.
    
    We have chosen the \texttt{Adam} optimizer\footnote{In the context of NN-training, the optimizer is the method that implements some algorithm for the optimization of the loss function, generally some form of stochastic gradient descent.} for the optimization process, which uses stochastic gradient descent with information about higher-order momenta (see \cite{adam} for details).
    
    \item The entire process is iterated for a fixed number of training cycles, known as epochs. During each epoch, we evaluate the error (loss) in our predicted field values for each equation to ensure convergence below a predefined error threshold.
\end{enumerate}

The described NN system has been implemented in the \texttt{Python} language through the open source \texttt{neurodiffeq} library \cite{neurodiffeq}, built on \texttt{PyTorch.}

The  hyper-parameters that we use include an initial learning rate of $0.001$ decreasing 1.5\% every five thousand epochs. All other parameters are kept at the {\tt PyTorch} default. For the solver network, we use 6 different networks with three hidden layers of 32 nodes with \texttt{tanh} activation. For the V-network, we employ 4 hidden layers with 16 nodes and \texttt{SiLU} activation. This architecture was determined through trial and error, exploring configurations that have previously proven effective for similar problems.

\subsection{Technical aspects of setup and training}
In this section, we describe in detail the main technical aspects of our method. The  setup is shown in \fig{fig:gaussian architecture}. As already mentioned, the NN-solver takes as inputs both the sampling of points of the independent variable $u$, and pairs of points $(T,S)$ along the thermodynamic curve, each of them corresponding to different boundary conditions on the functions $\tilde{\Sigma}$ and $\nu_A$. Let us describe some aspects of each input separately.

\subsubsection{Independent variable generator}

The independent variable $u$ runs from the AdS boundary ($u = 0$) to the location of the black brane horizon ($u = 1$). Since this is the independent variable of the differential equations and one of the inputs of our model, we first need to sample it in such a way that it covers the range of interest. To do so, we have decided to select $48$ points sampled using Chebyshev of the second kind nodes,\footnote{For a mathematical definition of Chebyshev polynomials, see \cite{chebyshev}.} that is, by placing a higher density of points near the endpoints of the interval. This results in a better resolution of  the value of the functions near both  boundaries.

We have also explored the possibility of sampling a different number of points, and using a different sampling technique. The main advantage of reducing the number of points in the 
$u$-direction is that the training process is  faster but, in general, we  obtain less precise solutions to the differential equations and, thus, a less precise reconstruction of the unknown potential. We have also explored the possibility of using an equally spaced distribution of points randomly displaced a small quantity (noise) at each training epoch. However, this sampling technique does not allow us to  resolve the boundary region as well as with Chebyshev sampling and  it results in slightly worse results. 

\subsubsection{Points on the equation of state}
\begin{figure}
    \centering
    \includegraphics[width=0.62\columnwidth]{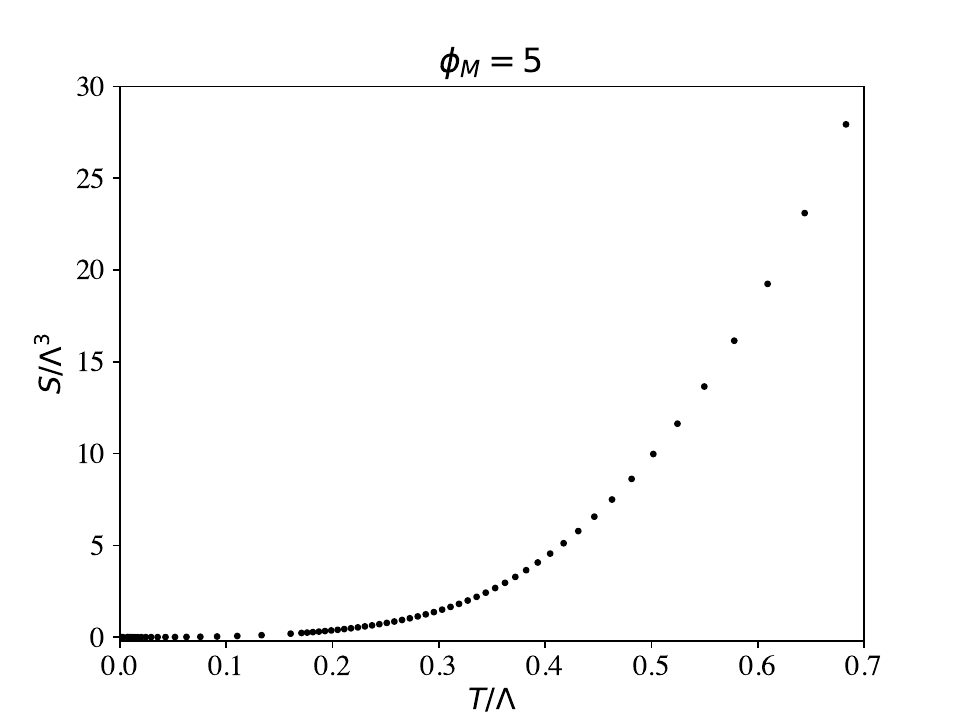}\\
    \includegraphics[width=0.62\columnwidth]{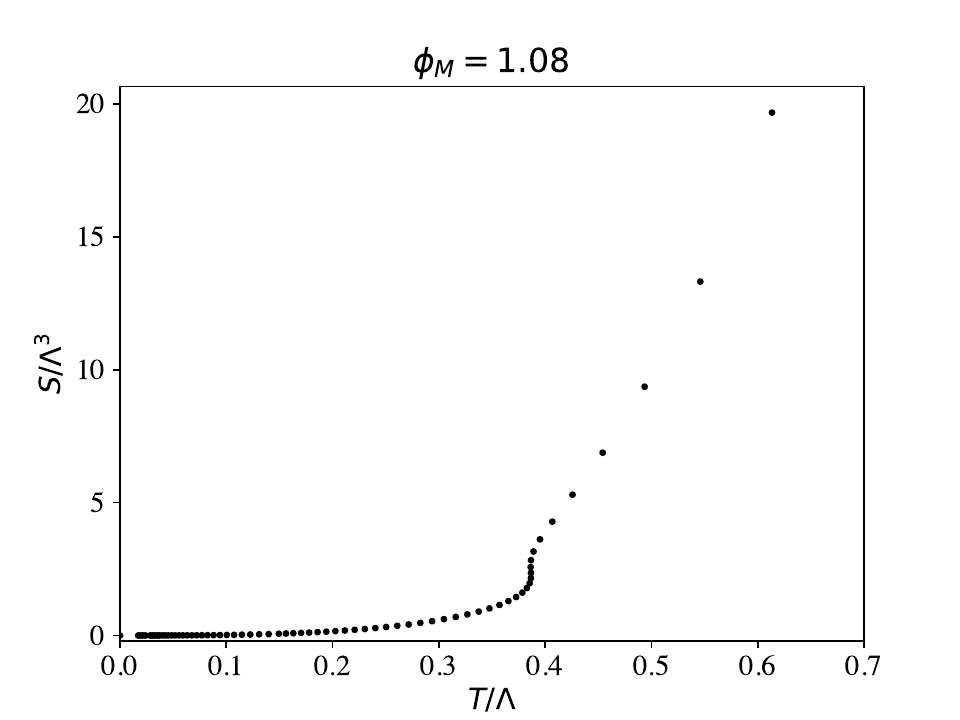}\\
    \centering
    \includegraphics[width=0.62\columnwidth]{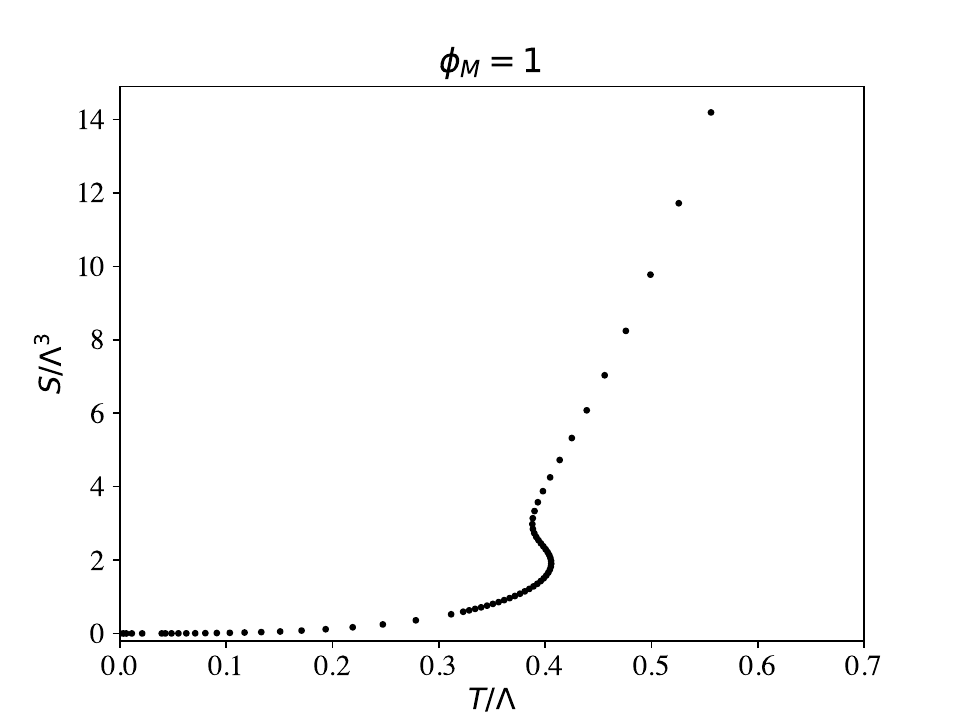}
    \caption{Sampling of the $S(T)$ curve for a crossover (top), a second-order transition (middle) and a first-order transition (bottom), in the range of temperatures $0$ to $0.7$. The points along the curves are sampled more densely around the phase transition or crossover ($T/\Lambda \approx 0.4$), as well as in the IR ($T/\Lambda  \approx 0$). }
    \label{fig: s(T) inputs}
\end{figure}

An important aspect regarding the inputs of the network is the number of points sampled along the $S(T)$ curve. On the one hand, a large number of points will lead to long training times. On the other hand, a small number of points will not provide the network with enough information to reconstruct the potential correctly. We have found that the number of points that provides a good balance between training time and precise recovered potentials is in between $65-70$, sampled for values of the temperatures ranging from $T/\Lambda=0$ to $T/\Lambda\simeq 0.6$. This range of temperatures translates into a range of entropies from $S/\Lambda^3=0$ to $S/\Lambda^3 \simeq 30$. The maximum value of the temperature is  chosen so that it is higher than the transition/crossover temperature, which is $T_c/\Lambda \simeq 0.4$, but not so high as to generate too large a hierarchy for the values of the entropy density.  Moreover, this sampling is not uniformly done for all values of the temperature. Instead, a higher density of points is chosen in the phase transition or crossover regions, as well as in the IR region of the curve --- see \fig{fig: s(T) inputs}. The reason for this is that the UV region of the potential is easily constrained, whereas more precision is needed in order to resolve correctly the parts of the potential that control the transition and the IR physics.

\subsubsection{Architecture and activations}
As illustrated  in \fig{fig:gaussian architecture}, the architectures of each of the two NNs are different. The first one, NN-Solver, is composed  of $6$ different nets, each one of them taking $u$ and the boundary conditions ($T,S$) as inputs and giving the solutions to the DEs as outputs. These nets have $3$ hidden layers with $32$ neurons each. The second NN, NN-V, acts as the scalar potential function $V(\phi)$, taking the solution  $\phi(u)$ coming from solving the DEs as input and giving the scalar potential as an output. This net is composed of $4$ hidden layers with $16$ neurons each. In summary, we have:
\begin{align*}
    [\mbox{NN-Solver }]_i:\,\,\,\,&\,[32,32,32]_i\,\,,\,\, \,\,\text{with}\,\,\, i=1,\dots,6 \,; \\
    \mbox{NN-V}\,\,\,:\,\,\,\, &\,[16,16,16,16] 
\end{align*}

We have chosen the standard hyperbolic tangent, $\tanh(x)$, for the activation functions of all neurons in the ODE nets {NN-Solver}. In contrast, we use the Sigmoid-weighter Linear Unit function  ($\texttt{SiLU}$, a.k.a.~Swish), $\texttt{SiLU}(x)=x(1+e^{-x})^{-1}$,
as the activation for the neurons in the hidden layers of the potential net {NN-V}. The fact that the SiLU function is not bounded from above, as opposed to $\tanh(x)$, allows for more general behavior of the output $V(\phi)$. This is a desirable property when predicting free-form functions that could a priori be extremely complicated, since they must reproduce the thermodynamics of a variety of different theories.

\subsubsection{Gaussian localization in {V-NN}} \label{subsubsec: gaussian localization}
We have developed a novel feature that we have dubbed ``Gaussian localization'' (GL), pictorially shown in \fig{fig:gaussian architecture}, that has improved the reconstructed potential. Here we will describe the method and in Sec.~\ref{disc} we will discuss its effect. The NN-V network is composed of $4$  layers whose response can be expressed as 

\begin{equation}\label{eq: linear layer response}
    \mathbf{h_i} \,=\, \texttt{SiLU} \left\{ \mathbf{h_{i-1}} \cdot \mathbf{W^{T}_{V,i}}  \right\} \,,
\end{equation}
 where $\mathbf{h_i}$ is the output vector of the $i$-th layer. In the case of $\mathbf{h_1}$ (response of the first hidden layer of NN-V), 
 $h_0=\phi$, where $\phi$ is the scalar field that is one of the outputs of the solver network. $\mathbf{W_{V,i}}$ is the weight matrix\footnote{It is customary to absorb the bias in the weight matrix by simply adding a column of ones.} (except $\mathbf{W_{V,0}^T}$ which is a vector).

This general framework is modified as follows by our GL: we multiply the output given by \eqref{eq: linear layer response} of the first hidden layer by a Gaussian function, resulting in
\begin{equation} 
    \mathbf{\tilde{h}_1} = \mathbf{h_1} \times \text{exp}\left[{-\left(\frac{\mathbf{h_1} - \bm{\mu}}{\bm{\sigma}}\right)^2}\right] \,,
\end{equation}
where $\mathbf{\tilde{h}_{1}}$ the output vector of neurons of the first layer after the localization. The difference is thus that the Gaussian multiplies the output of the neurons from the 1st layer, which is \eqref{eq: linear layer response} after passing it through the activation SiLU. The parameters $\bm{\mu}$ and $\bm{\sigma}$ are vectors of the means and standard deviations of the Gaussians corresponding to the localization for each neuron.  For the Gaussian widths we have chosen $\sigma = 0.1$, and we have let their centers, $\bm{\mu}$, be a learnable parameter. The result of this learning is that these centers are approximately evenly distributed throughout the range of values of $\phi$. As a consequence, each neuron in the first layer, and the subsequent paths that follow from its output, specialize in recovering a certain region of $V(\phi)$. To understand  the enhanced performance of this approach, consider how the network approximates the function $V(\phi)$ across the entire input domain. Given that we employ a fully connected network, a modification in the weights of the first hidden layer that may reduce the loss at a specific input value $\phi$ could potentially have a contrasting impact at another input value $\phi'$. It is imperative that the gradients of the loss with respect to the weights of the first layer exhibit a low correlation for different values of $\phi$. Essentially, the quantity $(\nabla_W L|_{\phi}) \cdot (\nabla_W L|_{\phi'}) $ should be minimized when $\phi$ and $\phi'$ are significantly distant. 
While there exist various methods to achieve this objective, a straightforward and efficient approach involves scaling the outputs of each node in the first layer by a Gaussian function, as previously outlined.
From a physics perspective, an alternative interpretation lies in the ``multi-scale'' and ``multi-entangled'' characteristics inherent in the gravitational aspect of the problem. The multi-scale aspect refers to the fact that the five-dimensional geometries can develop a large hierarchy of scales between the near-boundary and the near-horizon regions. We will come back to this point in Sec.~\ref{disc}. Multi-entanglement  refers to the fact that the value of the potential at a certain value $\phi=\phi_0$ affects the thermodynamic properties of \emph{all} the black brane geometries with horizons such that \mbox{$\phi_H \geq \phi_0$}. Our results suggest that a NN in which each neuron specialises in a certain region or scale is better suited for solving a multi-scale and multi-entangled problem. 

We have also examined the effect of GL on the network NN-Solver that we have used to solve the DEs. Note that since this model  has three inputs, $u$, $T$ and $S$, three independent localizations are needed, one for each of these variables. 
Localizing in a multi-dimensional space presents significant challenges due to the extensive parameterization. Currently, the methods that we have employed for localizing in dimensions such as $u$, $T$, and $S$ have not yielded the desired improvements in training. In fact, they have hindered progress rather than facilitating it. As we continue our efforts, we are exploring alternative approaches and refining our strategies to enhance the localization process and improve overall training outcomes. For now, we have decided not to include it in this work.

We have also explored the possibility of letting the network learn the values of the Gaussian widths $\sigma_j$ in NN-V. However, this leads to only a couple neurons in the first layer controlling the whole output of the layer, resulting in an augmentation of free parameters that does not improve the fit. 

\subsection{Numerical tests}
\label{numtest}
Here we  perform two numerical checks of the model. 
First, we verify that the NN is solving the DEs correctly. 
For this purpose, rather than reconstructing the potential, we fix it to be one of the potentials in \fig{fig: V for different phiM}. Then  we ask the NN-Solver to solve the direct problem with this potential, namely, to solve the system of ODEs \eqref{Ealpha} for three different points along the $S(T)$-curve. We then compare these solutions with their ``theoretical'' values obtained with a traditional method \cite{Bea:2018whf}. For each function, we define the error as the square of the difference between the function produced by the NN and the theoretical function. Since these functions  take values of order unity in most of their domain (see Figs.~\ref{fig:solutionsASigmaPhiM1} and \ref{fig:solutionsASigmaPhiphiM5}), the square root of this error is a good measure of the relative error between the theoretical and the NN-predicted functions.  

The three different pairs $(T,S)$ that we have given to the NN correspond to different boundary conditions of the functions $\tilde{\Sigma}(u)$, $\tilde{A}(u)$ and $\phi(u)$ at the horizon. We have chosen these points to probe the quality of the solutions in three different regions of the $S(T)$-curve. The first one (high temperature) is chosen to be around $T/\Lambda\simeq 0.5$ and  tests the solution near the AdS boundary. The second one  is placed around $T/\Lambda \simeq 0.4$ (intermediate  temperature) and it  probes the most ``non-conformal'' region of the boundary theory. The third point is chosen to be near $T/\Lambda\simeq 0.25$ (low temperature) and it probes the solutions near the IR fixed point. We have performed this test for two different theories characterized by $\phi_M = 5$, which possesses a crossover, and $\phi_M = 1$, which corresponds to a first-order phase transition. For the first case and for $1\cdot 10^{6}$ epochs, the highest error is found for the low-temperature solution and it has a value of approximately $1.4 \cdot 10^{-4}$.  This maximum error decreases substantially when increasing the number of epochs to $5\cdot 10^{6}$, giving a final value of $7\cdot 10^{-6}$. For the case $\phi_M = 1.0$ we have also found the maximum error for the low-temperature solution, with a value of $1.0\cdot 10^{-3}$ when the model is trained for $1\cdot 10^{6}$ epochs. As before, if we increase the number of epochs up to $5\cdot 10^{6}$ then this error decreases down to $2.5\cdot 10^{-4}$. The main conclusion of this test is that the {NN-Solver}  is able to correctly solve the system of ODEs with sufficiently high precision when the  function $V(\phi)$ is specified. The fact that the largest errors occur for the low-temperature solutions is consistent with the multiscale, multinentangled nature of the problem discussed in \ref{subsubsec: gaussian localization}, since these solutions depend on the value of the potential over a longer range of $\phi$ values.

We have performed several tests regarding the choice of the sampling technique and the number of points of the independent variable. As we have mentioned, for the sampling techniques we have used Chebyshev nodes and equally-spaced points with noise. For each  of these methods we have tested the model's  performance for $32$ and $48$ points, for the $S(T)$-curve  corresponding to the $\phi_M  = 5$ case. In general, we have observed that the Chebyshev sampling yields better results, in the sense that we have obtained losses of the order of $10^{-6}$ using equally-spaced noisy points, and losses of $10^{-7}$ using Chebyshev nodes. Moreover, we have observed that this last method is much more controlled and consistent. Indeed,  in both cases the learning rate  needs to be adjusted manually at some point, but the random component of the noisy equally spaced points makes if very difficult to estimate \textit{when} this should be done. In contrast, for the Chebyshev technique it is very clear that this must be done when the loss starts flattening (see \fig{fig: loss}).

Regarding the number of points, we have observed that the overall loss can be lower for 32 points than for 48.  For example, the lowest loss we found was $8.18 \cdot 10^{-8}$, which occurred for the $\phi_M=5$ model with Chebyshev nodes and $32$ points in the $u$-direction. In contrast, with 48 points we obtained a loss of $1.16 \cdot 10^{-7}$. Nevertheless, we have also observed that the recovered potential captures the UV behaviour near $\phi = 0$ more precisely with 48 points. 
For this reason, we decided to use $48$ points in the $u$-direction in the final algorithm.

The training process of the NN is subject to a certain stochasticity due both to the nature of the stochastic gradient descent and to the random distribution of the initial weights and biases. In order to estimate the magnitude of this uncertainty,  we have  trained the same model multiple times for a given number of epochs. The small dispersion of the results gives an idea of the robustness of our method. In addition, we have found that this is a more efficient way of obtaining a good solution, as compared to training the model only once for much larger number of epochs. In summary, we performed 10 different runs of order $\sim 10^6$ epochs for each model and, in some cases, we continued the training for the best run.

\subsection{Computational considerations} 
The total number of free parameters of our model (weights, biases and Gaussian parameters) is $14519$, of which $881$ belong to $W_V$ and $13638$ belong to $W_D$. This gives us approximately $1.5\cdot 10^{4}$ adjustable parameters to train our model to solve the differential equations and to reconstruct the potential.

Training and discovery of a solution with sub-per cent precision in the potential and per cent precision  in the recovered equation of state requires about 8-16 hours and a couple of million epochs in a dedicated NVidia A40 GPU. As explained above, we performed ten runs for each $(T,S)$-curve. 
The solutions discovered could be used for other problems using transfer learning and will be added to the \href{http://dev.neurodiff.io}{neurodiffHUB} repository. These solutions, together with our code, will  be made publicly available in the near future.

\section{Results}
\label{results}
 We test our NN algorithm using boundary conditions associated to the $S(T)$-curves in \fig{SofT}. Different values of $\phi_{M}$ correspond to different theories with a crossover, a second- or a 
first-order phase transition. These differences are encoded in the different shapes of the thermodynamic curves $S(T)$, as can be seen in Fig.~\ref{SofT}. As explained in Secs.~\ref{holomodel} and \ref{sec: methodology}, we find solutions for the dependent variables $\tilde{\Sigma}(u)$, $\tilde{A}(u)$ and $\phi(u)$, one for each boundary condition, 
i.e.~one for each point in $S(T)$. These are solutions to the  system of ODEs \eqref{Ealpha}, which depends on the unknown potential $V(\phi)$. This potential is the same for, and is informed about, all the boundary conditions, and we find it  (necessarily) at the same time as the rest of the functions.  We obtain the following results.

\subsection{Loss function}
\begin{figure}  
    \centering
    \includegraphics[width=0.8\textwidth]{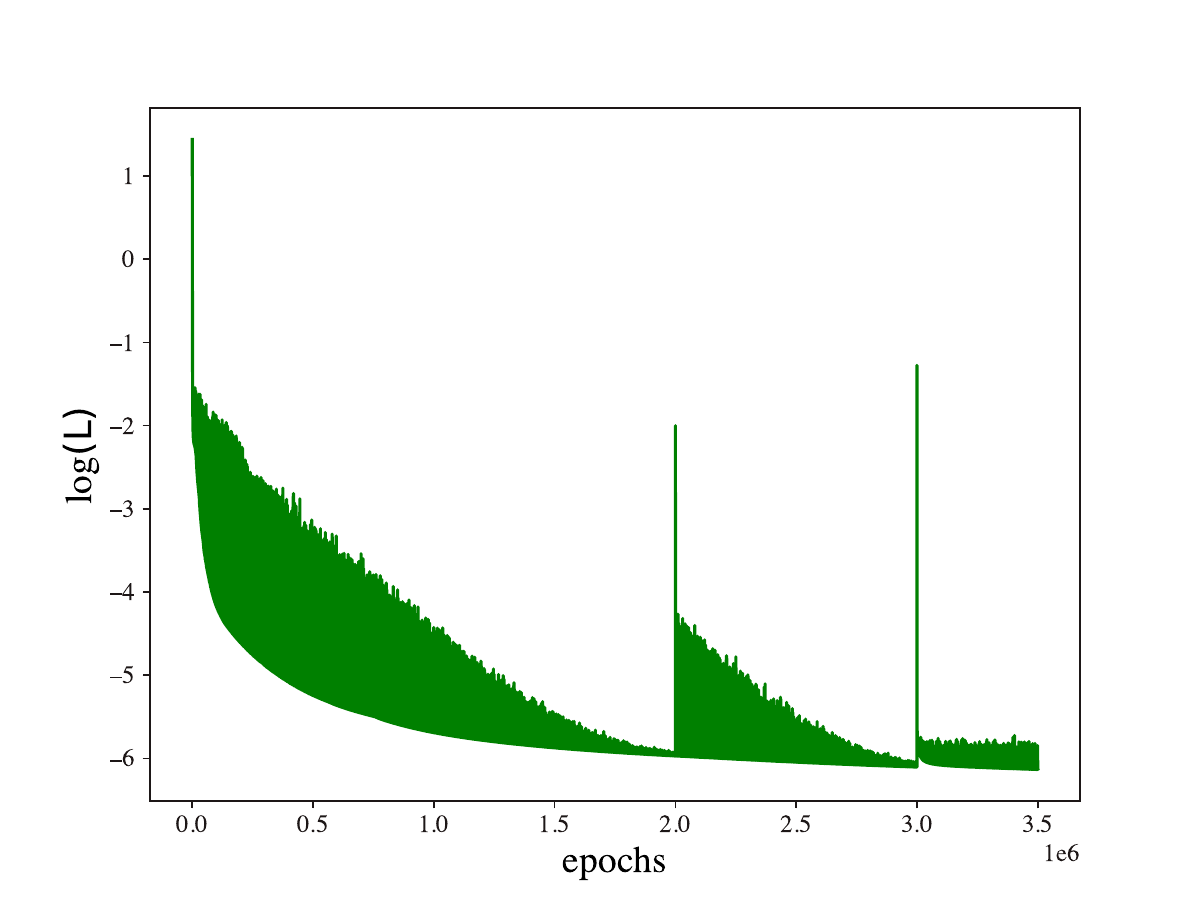}
    \caption{Loss function (squared residual from our ODEs) in log scale as a function of the epoch, for the best run of the first order transition case ($\phi_M=1$). This NN has been trained for 3.5  million epochs. The minimum loss achieved is $\approx7.4\cdot 10^{-7}$.} 
    \label{fig: loss}
\end{figure}

Our method is based on minimizing the loss function \eqref{loss}.   \fig{fig: loss}  shows the typical form of this function, in this case  for the best run for the \mbox{$\phi_M=1$} model, which possesses  a first-order phase transition case. We see how $L$ greatly decreases in the first half a million epochs and keeps doing so at an increasingly slower rate for the next $2.5$ million epochs. It is interesting to note that, for our choice of hyperparameters, the $1.5\%$ decrease in the learning rate every $5000$ epochs (see Sec.~\ref{subsubsec: NN setup and procedure}) makes the training more efficient only up to a certain point, after which the learning rate becomes too small compared to the value of the loss, and the loss flattens. At this point, we say that the NN is ``trapped'' in some local minimum. To continue with efficient training, one must increase the learning rate by hand. This can be seen at 2 and 3 million epochs in  \fig{fig: loss}. Automatization of this procedure is part of our future work that could incorporate techniques such as annealing.

\subsection{Recovered potential and equation of state}

\begin{figure}
    \centering
    \includegraphics[width=0.92\columnwidth]{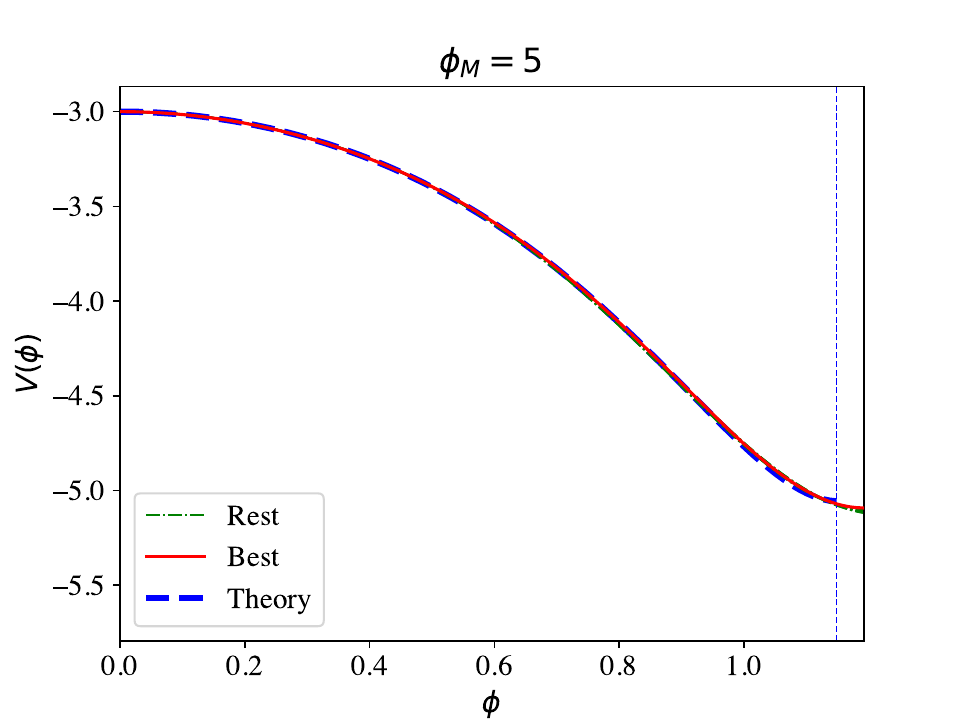} 
     \includegraphics[width=0.92\columnwidth]{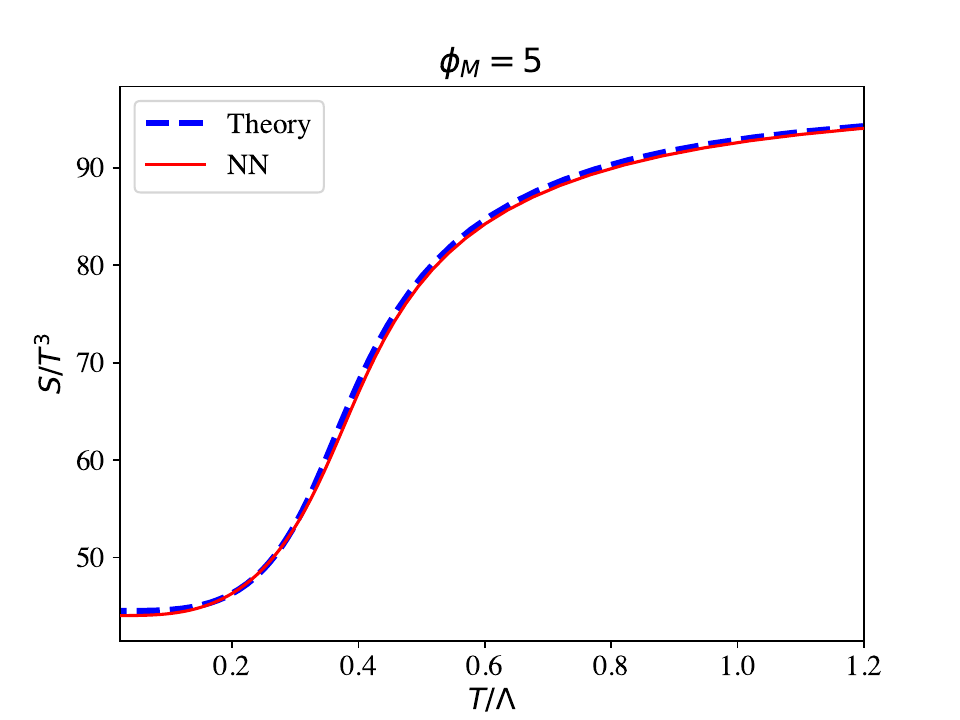} 
    \caption{(Top) Best scalar field potential predicted by {NN-V} (solid red) compared to the theoretical potential (dashed blue) for a theory with a crossover. The plot extends up to the position of the predicted minimum of the potential, while the vertical line (dashed blue) indicates  the position of the theoretical minimum. (Bottom) Recovered equation of state (solid red), obtained by  solving the direct problem with the recovered potential, compared to the theoretical one (dashed blue).}
    \label{ComparisonCrossover}
\end{figure}

\begin{figure}
    \centering
    \includegraphics[width=0.92\columnwidth]{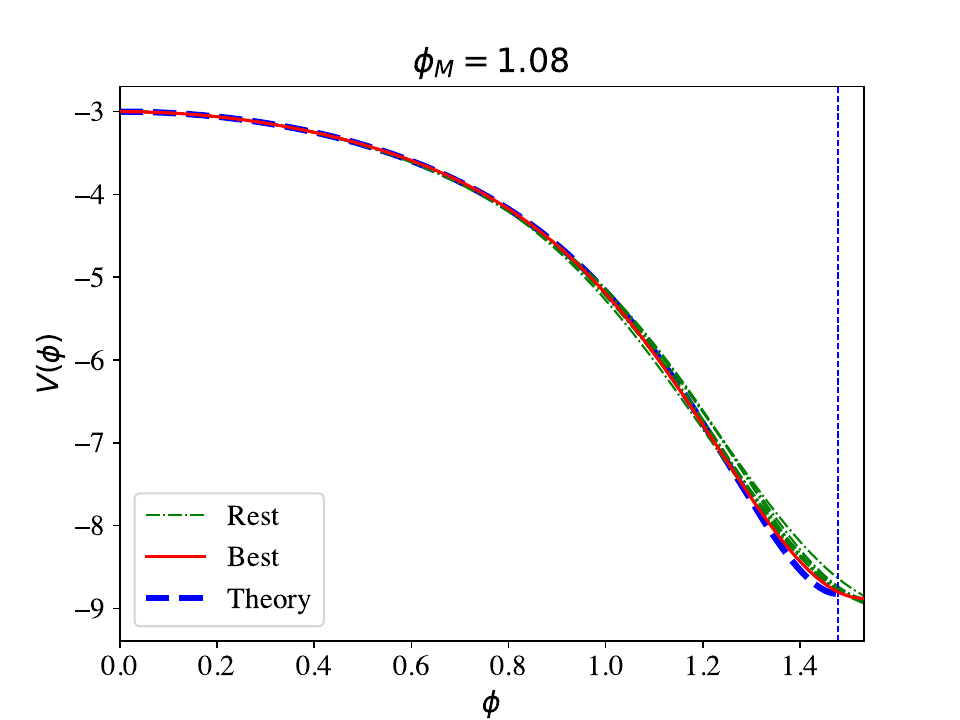} 
     \includegraphics[width=0.92\columnwidth]{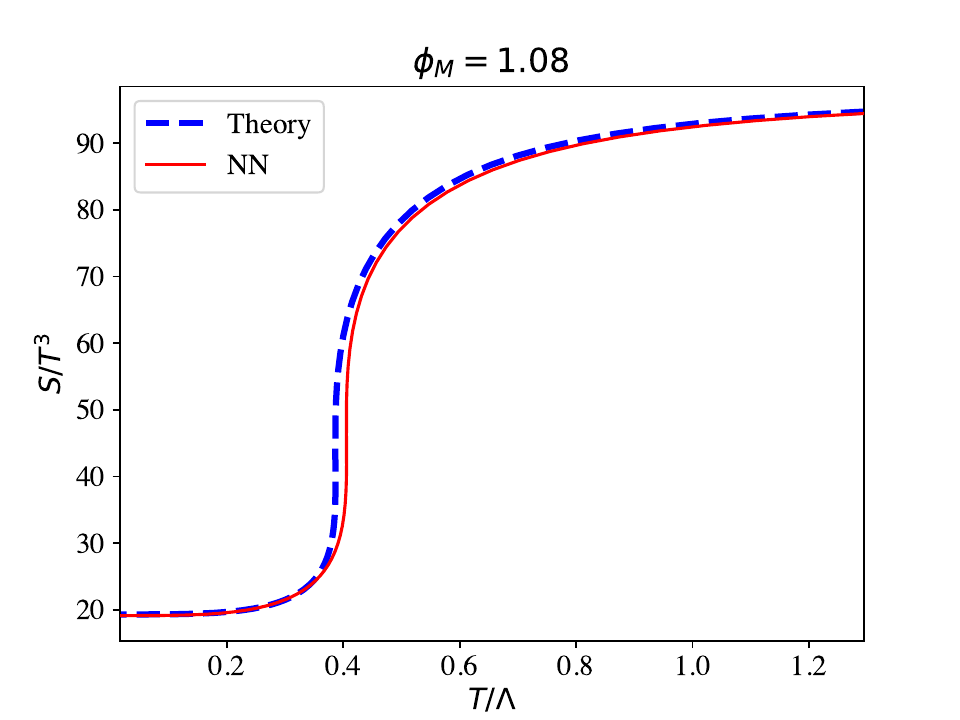} 
    \caption{(Top) Best scalar field potential predicted by {NN-V} (solid red) compared to the theoretical potential (dashed blue) for a theory with a second-order phase transition. The plot extends up to the position of the predicted minimum of the potential, while the vertical line (dashed blue) indicates  the position of the theoretical minimum. (Bottom) Recovered equation of state (solid red), obtained by  solving the direct problem with the recovered potential, compared to the theoretical one (dashed blue).}
    \label{Comparison2ndOrder}
\end{figure}

\begin{figure}
    \centering
    \includegraphics[width=0.92\columnwidth]{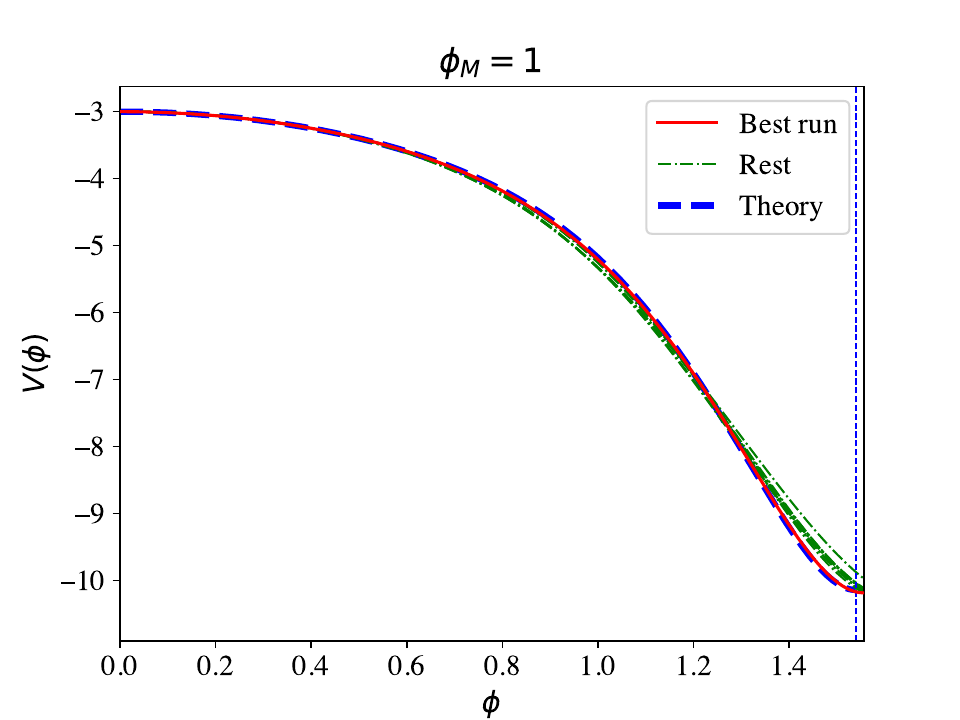} 
     \includegraphics[width=0.92\columnwidth]{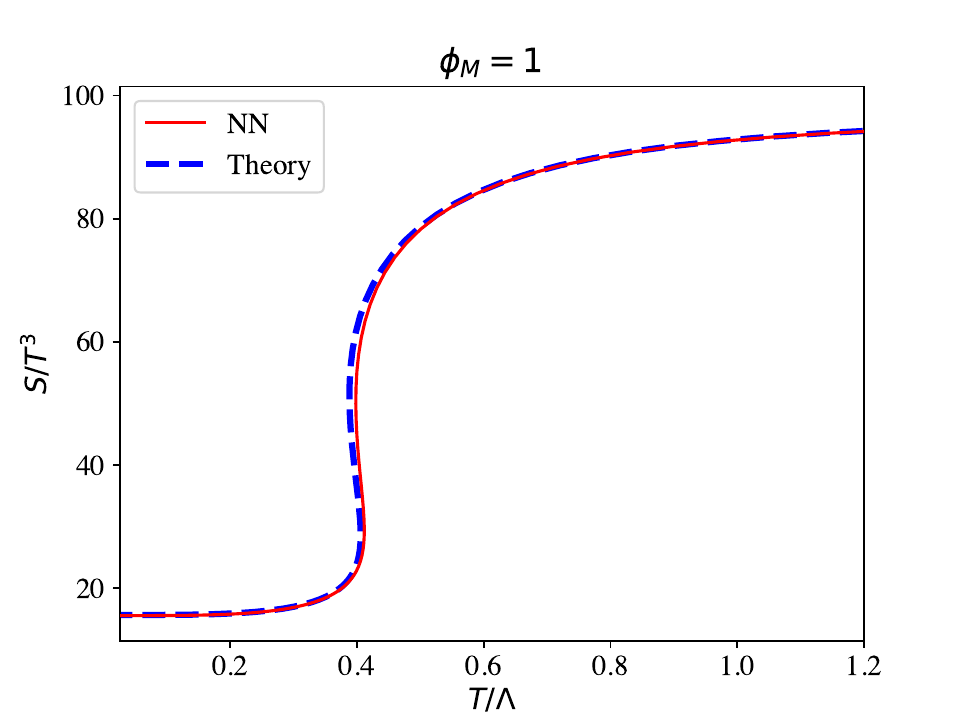} 
    \caption{(Top) Best scalar field potential predicted by {NN-V} (solid red) compared to the theoretical potential (dashed blue) for a theory with a first-order phase transition. The plot extends up to the position of the predicted minimum of the potential, while the vertical line (dashed blue) indicates  the position of the theoretical minimum. (Bottom) Recovered equation of state (solid red), obtained by  solving the direct problem with the recovered potential, compared to the theoretical one (dashed blue).}
    \label{Comparison1stOrder}
\end{figure}

\begin{figure}
    \centering
    \includegraphics[width=0.85\columnwidth]{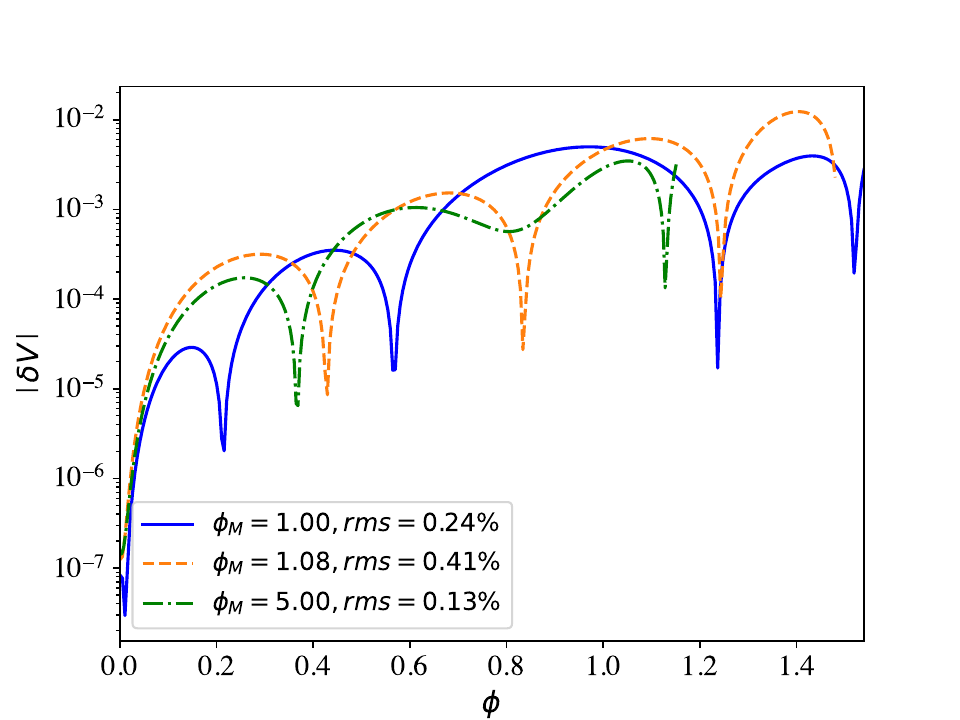} 
    \includegraphics[width = 0.85 \columnwidth]{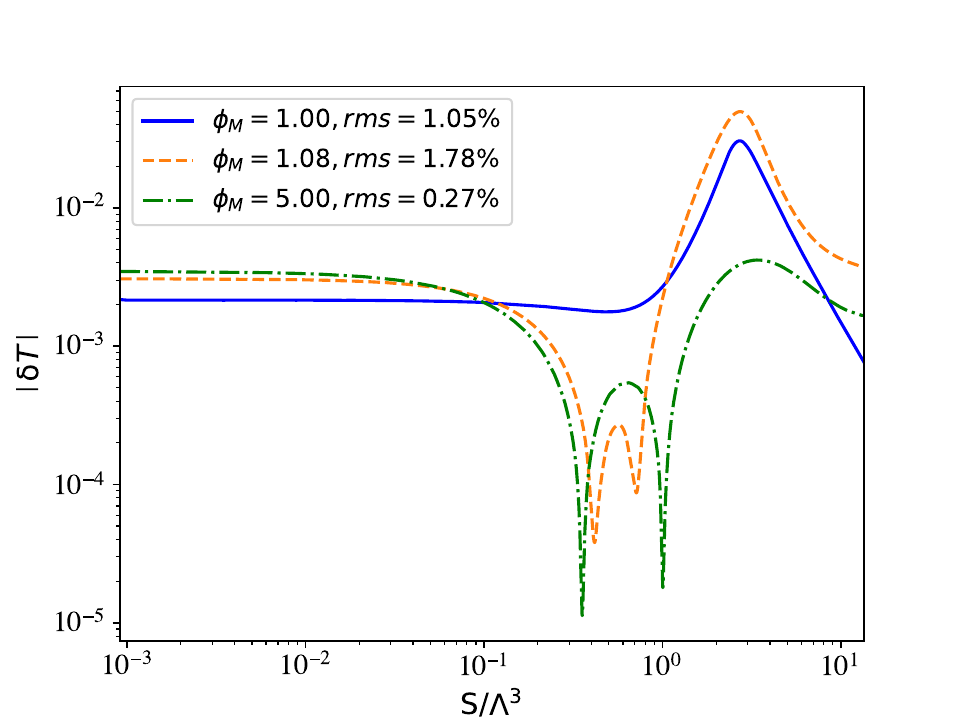} 
    \caption{Relative error for the potential (top) and for the recovered equation of state (bottom) for the best run. These quantities have been computed using the expressions (\ref{eq: relative error V}) and  (\ref{eq: relative error T(S)}). In the legend, we show the rms for each value of $\phi_M$. Note that $|\delta V|$ (top) extends over a longer range of $\phi$ for the cases with lower $\phi_M$. This is because the position of the minimum of the scalar  potential $V(\phi)$ is at a larger $\phi$ for cases with smaller $\phi_M$ (see \fig{fig: V for different phiM}). Also, note that $|\delta T|$ (bottom) tends to be a constant for $S/\Lambda^3\rightarrow 0$. The small value of this constant indicates that the PINN reconstructs the IR CFT with precision.}
    \label{fig:residuals}
\end{figure}

In the top panels of Figs.~\ref{ComparisonCrossover}, \ref{Comparison2ndOrder} and \ref{Comparison1stOrder} we show the potentials produced by our method when supplied with the $S(T)$ curves corresponding to theories with a crossover, a second- and a first-order phase transition, respectively.  For each case, we run 10 realizations, and we select the best run as the one with the lowest value of the loss function. The resulting potentials for these best runs are shown as solid red curves in the figures. The remaining 9 runs are shown as dash-dotted green curves. The small dispersion between these curves illustrates the robustness of our method. Finally, the theoretical potentials \eqref{Vtheory} are shown as dashed blue curves. As is clear from the curves, the recovered potential is in general in excellent agreement with the exact one. For reasons discussed in Sec.~\ref{disc}, the precision of the recovery degrades as we move from a crossover to a second- to a first-order phase transition. We quantify this with the relative error
\begin{equation}
    \label{eq: relative error V}
    \left|\delta V(\phi) \right|=\, \left| \frac{V_\text{theory}(\phi) - V_\text{PINN}(\phi)}{V_\text{theory}(\phi)}\right| \,.
\end{equation} 
 We plot this error in \fig{fig:residuals}(top), where we  see that it is below $\sim 1\%$ for all values of $\phi$ for all theories. In this figure we also list the relative root mean square\footnote{This is defined as $\text{rms}=\sqrt{\frac{1}{n}\sum_{i=1}^n x_i^2}$ with  $x_i$ is the relative error defined in equations \eqref{eq: relative error V} and \eqref{eq: relative error T(S)}.} (rms), which is below $0.5\%$.

As explained in Sec.~\ref{holomodel}, the test for accuracy  shown in \fig{fig:residuals}(top) can only be done if one \textit{knows} the theoretical potential $V_\text{theory}(\phi)$. This implies having previously solved the ``direct problem'' to obtain $S_\text{theory}(T)$ from 
$V_\text{theory}(\phi)$.  However, if the input curve $S_\text{input}(T)$ has been obtained in a different manner than solving the direct problem (from the lattice, from observations, etc.), then one would not have the theoretical potential to compare $V_\text{PINN}(\phi)$ with. 

We thus turn to a more stringent test for accuracy. We take the best-recovered potential $V_\text{PINN}(\phi)$ and use it to solve the direct problem. This produces a curve $S_\text{PINN}(T)$ that we then compare with $S_\text{input}(T)$. The results are shown in the bottom panels of Figs.~\ref{ComparisonCrossover}, \ref{Comparison2ndOrder} and \ref{Comparison1stOrder}, respectively. We plot $S/T^3$ instead of $S$ because this exhibits the UV and IR behaviors more clearly. As in the case of the potential, we quantify the relative error as follows.
\begin{equation}\label{eq: relative error T(S)} 
  \left|\delta T(S)\right|= \left|\frac{T_\text{input}(S)-T_\text{PINN}(S)}{T_\text{input}(S)}\right| \,.
\end{equation}
We use $T(S)$ instead of $S(T)$ because the latter is  multi-valued, whereas the former is not. This error is shown in \fig{fig:residuals}(bottom). The phase transition is replicated fairly accurately. For example, the largest relative error in the recovered $S(T)$ occurs in the case of a second-order phase transition illustrated in \fig{Comparison2ndOrder}, and it is $\sim 5\%$. We also note that not only the PINN ``discovers'' the presence of an IR fixed point, but it predicts the number of IR degrees of freedom with remarkable precision. This is measured by the small error, of order $\sim 0.2\%$, in $\delta T( S=0)$, as illustrated  in \fig{fig:residuals}(bottom). In addition, the rms, which is an ``integrated'' measure of the overall error, is below $\sim1\%$ for theories with a crossover, and around $\sim1\%$ for  theories with a phase transition. We thus conclude that the entire RG  flow is correctly reconstructed.

\subsection{Solutions to the ODEs}
\begin{figure} 
    \centering
    \includegraphics[width = \columnwidth]{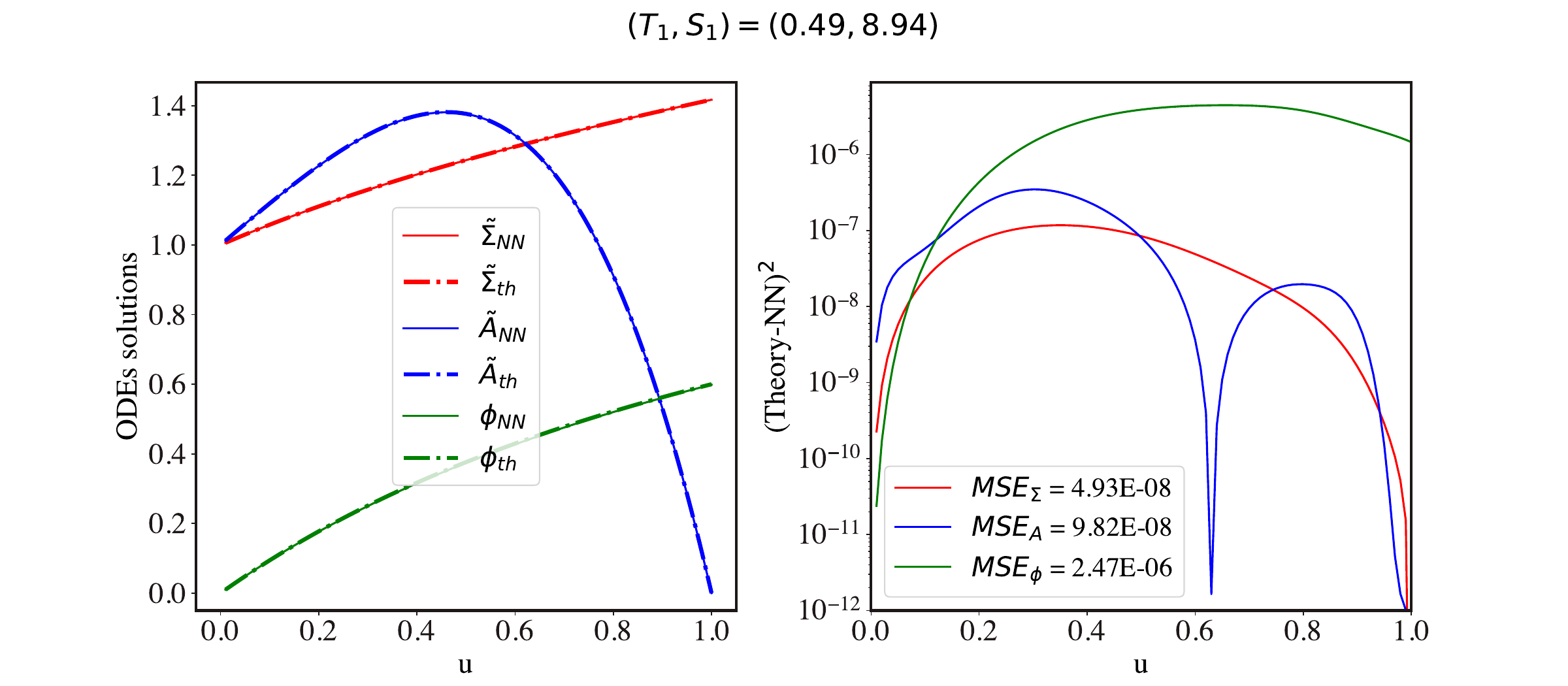}\\
    \includegraphics[width = \columnwidth]{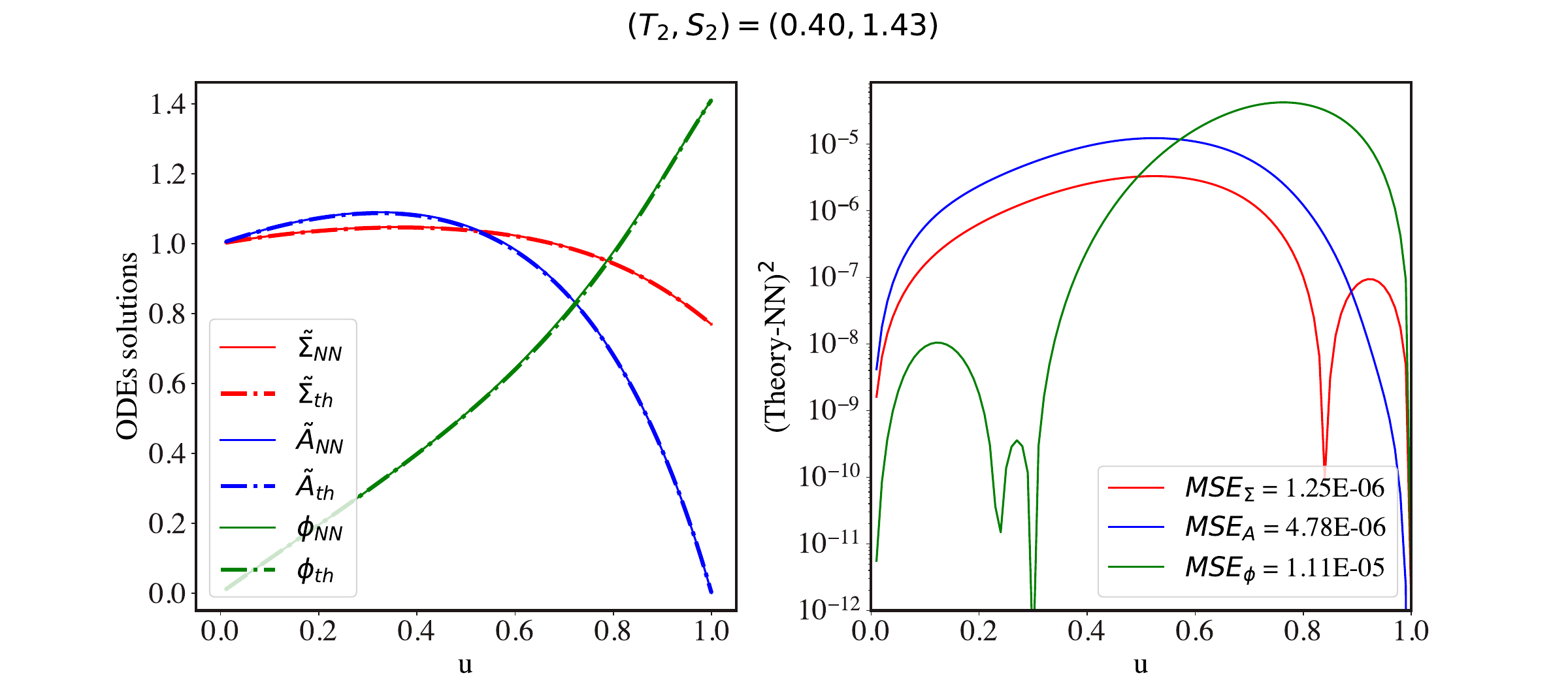}\\
    \includegraphics[width = \columnwidth]{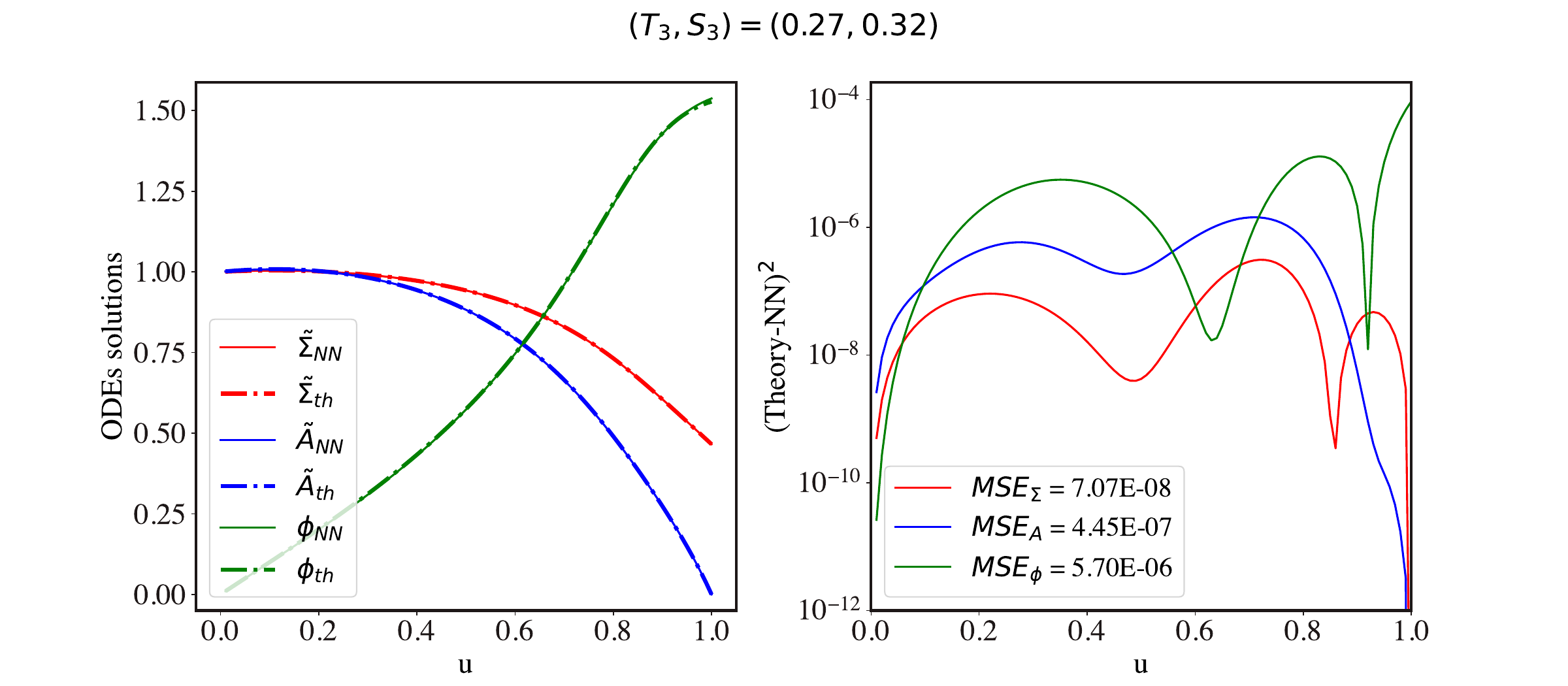}
    \caption{(Left) Comparison between the theoretical (dashed curves) and the PINN (solid curves) solutions for the theory with  $\phi_M = 1$. (Right) Squared differences between the theoretical and the PINN solutions.}
    \label{fig:solutionsASigmaPhiM1}
\end{figure}

\begin{figure} 
    \centering
    \includegraphics[width = \columnwidth]{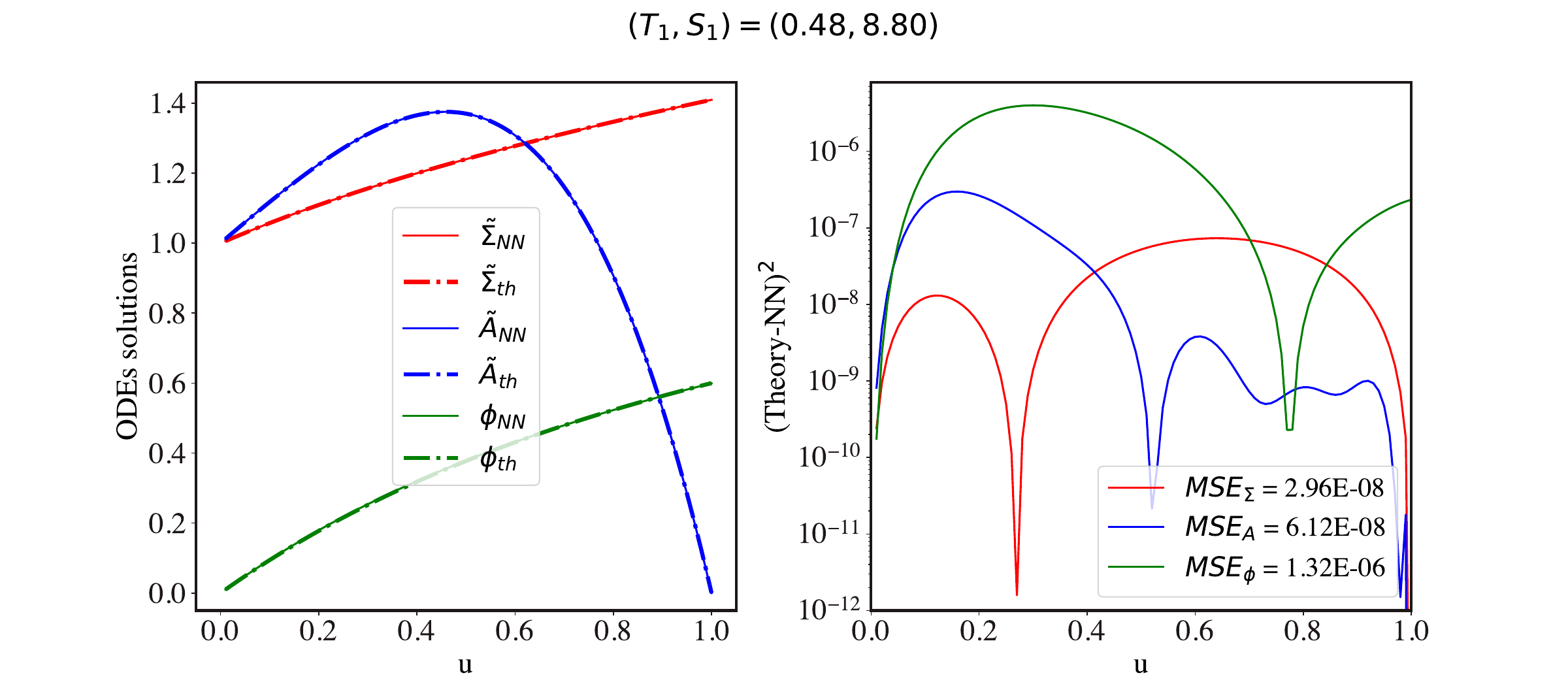}\\
    \includegraphics[width = \columnwidth]{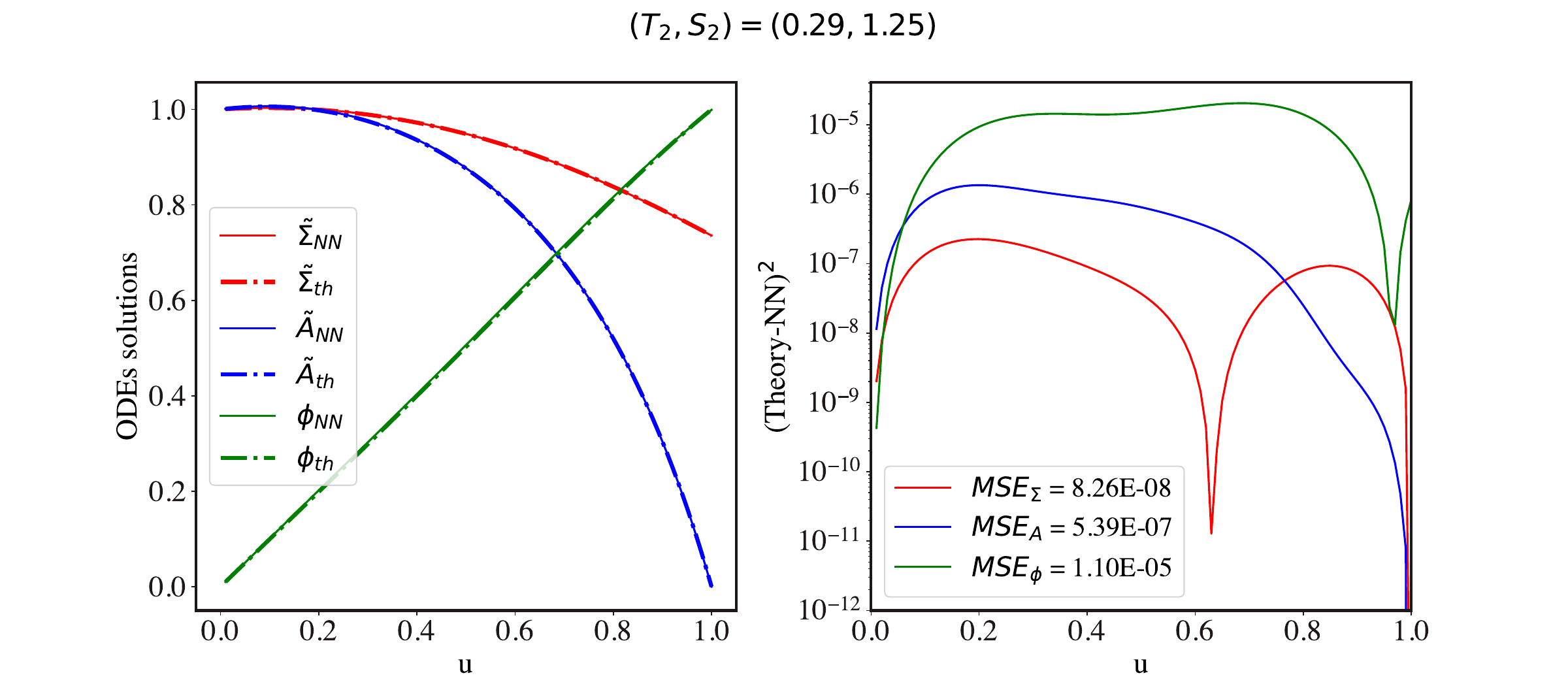}\\
    \includegraphics[width = \columnwidth]{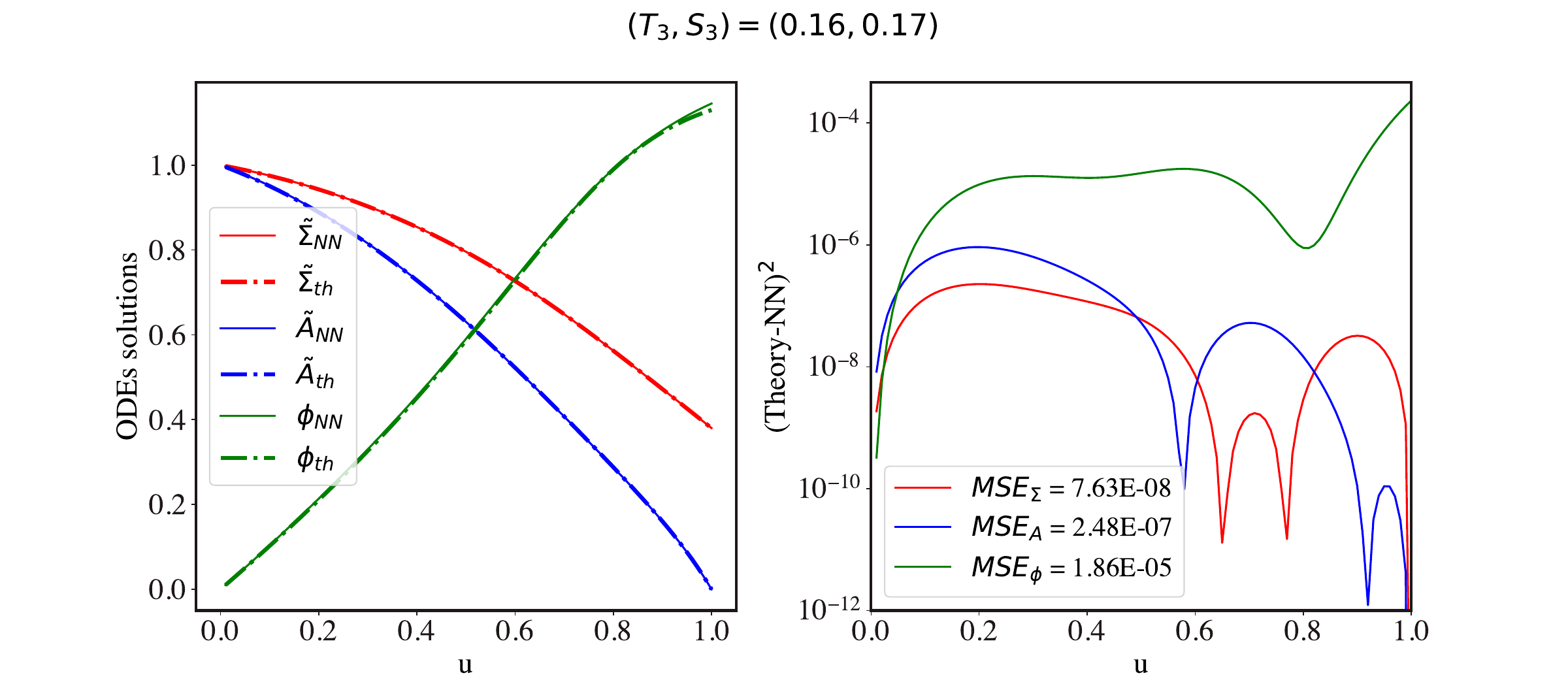}
    \caption{(Left) Comparison between the theoretical (dashed curves) and the PINN (solid curves) solutions for the theory with  $\phi_M = 5$. (Right) Squared differences between the theoretical and the PINN solutions.}
    \label{fig:solutionsASigmaPhiphiM5}
\end{figure}

The correct reconstruction of the potential requires solving the EKG accurately. Here we show the results for these solutions, namely, for the functions $\tilde{\Sigma}(u), \tilde{A}(u)$ and $\phi(u)$. There is one solution for each boundary condition, i.e.~for each point in $S(T)$. We show these solutions for the theories  with $\phi_M=1$ (\fig{fig:solutionsASigmaPhiM1}) and $\phi_M=5$ (\fig{fig:solutionsASigmaPhiphiM5}). For each theory, we show the solutions for 3 different points on the $S(T)$ curve that correspond to high, intermediate and low temperature in the language of Sec.~\ref{numtest}. 
 We can see an excellent agreement for both theories  between the  ``theoretical'' solutions (dash-dotted curves) and the NN-predicted ones (solid curves), with a mean squared error\footnote{This is defined as $\text{MSE} = \frac{1}{n}\sum_{i=1}^n \left(x_i^{(\text{th})} - x_i^{(\text{NN})}\right)^2$.} (MSE) of order $10^{-6} - 10^{-5}$.  
 Note that, since the functions $\tilde{\Sigma}(u), \tilde{A}(u)$ and $\phi(u)$ take values of order unity in most of their domain, the square root of the MSE is a good measure of the relative error between the theoretical and the NN-reconstructed functions.

\section{Discussion}
\label{disc}
Holography maps the quantum properties of a gauge theory in four dimensions to the classical properties of a gravitational theory in five dimensions. If the gravitational theory is known, the equation of state of the dual gauge theory can be found by solving a direct problem, i.e.~by finding all the black hole solutions of the gravitational theory. The inverse problem, namely to determine a gravitational theory that gives rise to a prescribed equation of state, is much more challenging. We have shown that this problem can be solved using Physics Informed Neural Networks. The resulting algorithm reconstructs not just a specific black hole solution but the gravitational theory itself. We have illustrated the method in a simple setup in which the gravitational theory is completely specified by one function, the potential $V(\phi)$ for a scalar field $\phi$. However, we expect that the method can be generalized to gravitational theories with a general field content, as well as to inverse problems outside the holographic context involving highly non-linear partial differential equations.

Figs.~\ref{ComparisonCrossover}, \ref{Comparison2ndOrder} and \ref{Comparison1stOrder} illustrate the reconstructed potentials, as well as the equations of state that they  yield upon solving the direct problem with them. The quality of the results is  quantified  in \fig{fig:residuals}, where we see that the relative error is at the sub-per cent level for the potential and of a few per cent for the equation of state. 
To obtain these results we provided the PINN with the boundary conditions \eqref{Vconditions} for the potential. These are encoded in the high-temperature behaviour \eqref{leadingStheory} of the equation of state and correspond to the fact that the gauge theory is a CFT deformed by a relevant operator of dimension \eqref{delta}.  

\begin{figure}
    \centering
    \includegraphics[width=0.62\textwidth]{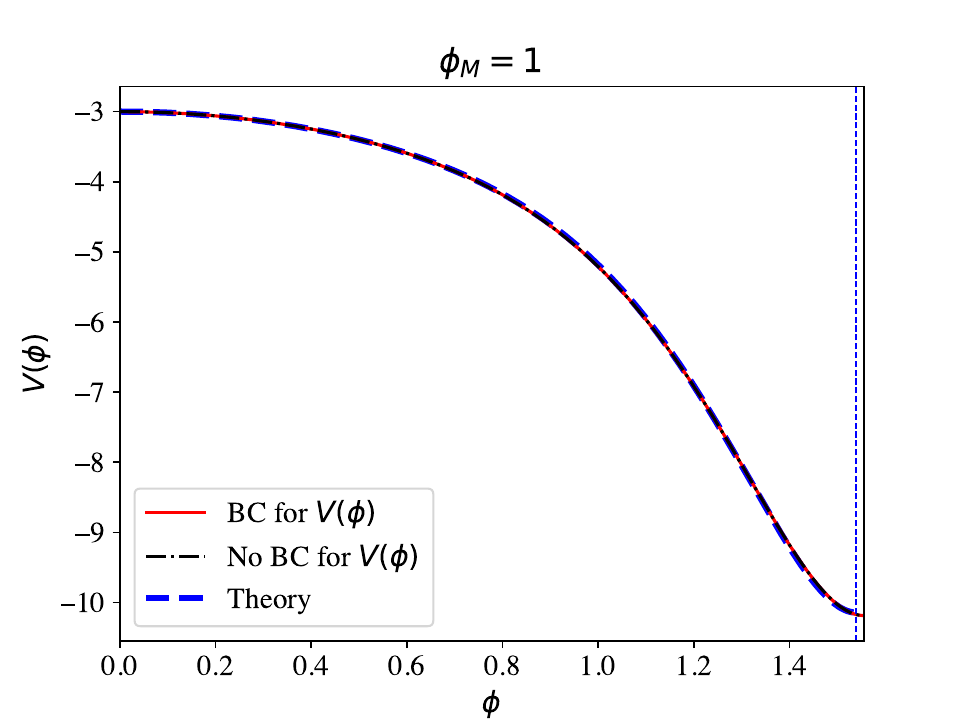}
    \includegraphics[width=0.62\textwidth]{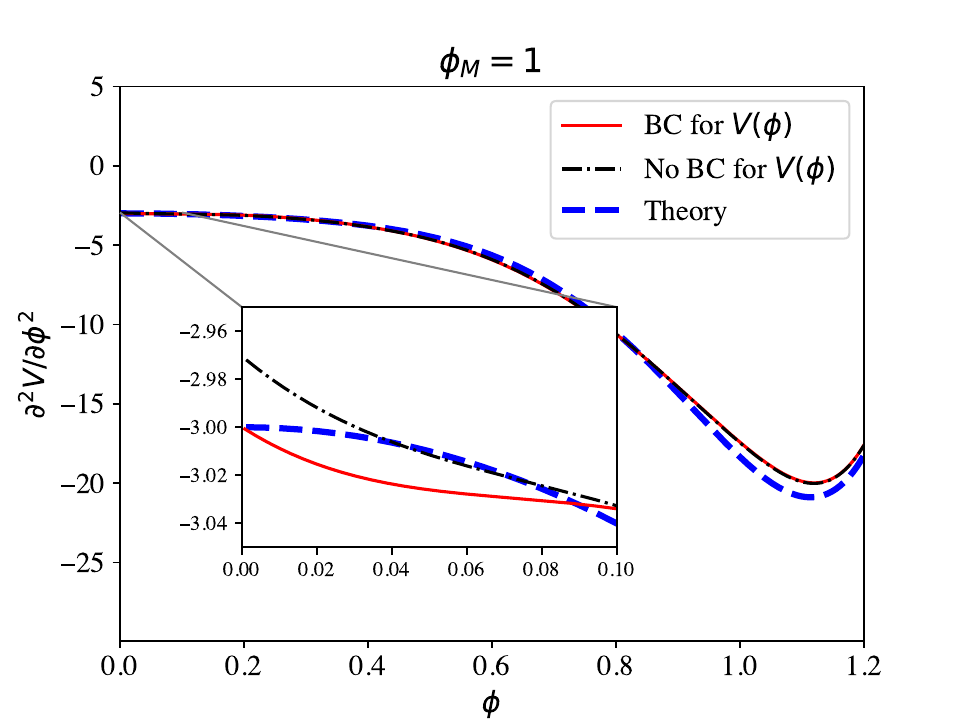}
    \includegraphics[width = 0.62\textwidth]{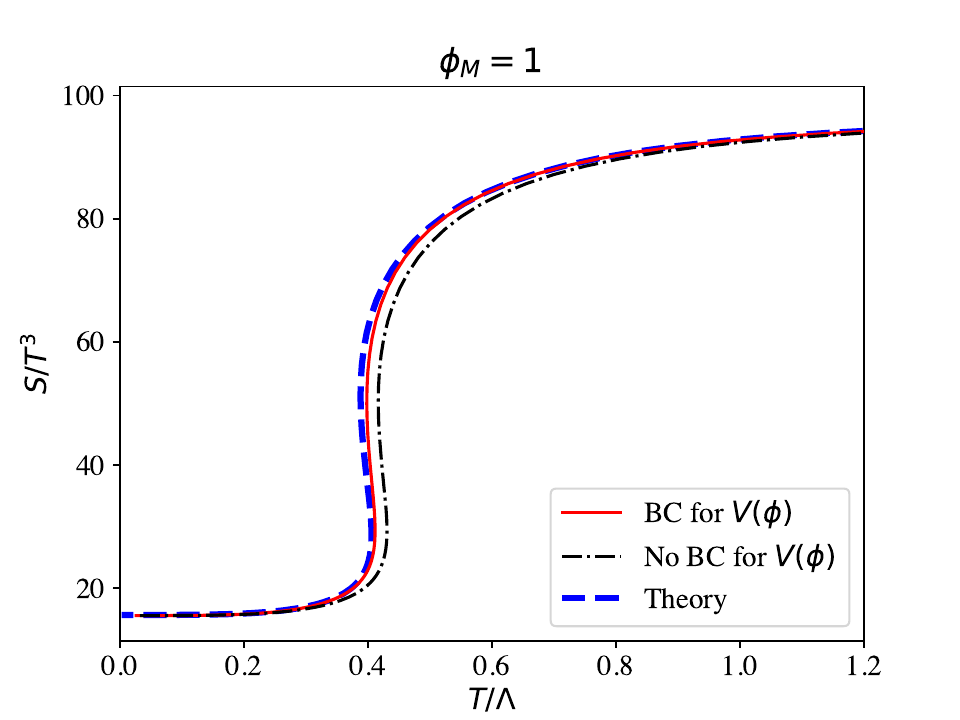} 
    \caption{Comparison between the reconstructed potentials (top), their second derivatives (middle) and the corresponding equations of state (bottom) when the PINN is or is not provided with the boundary conditions \eqref{Vconditions} for the potential.}
    \label{noBCforV}
\end{figure}

We have also trained the PINN without providing it with the boundary conditions \eqref{Vconditions} for the potential. In \fig{noBCforV} we compare the results to those obtained by providing the PINN with \eqref{Vconditions} for the theory with $\phi_M=1$. 
We see that, when the PINN is not provided 
with \eqref{Vconditions}, it is still able to recover the entire potential with high precision. In particular, it is able to identify the presence of a maximum in $V(\phi)$ at $\phi=0$, and to estimate $V(0)$ and $V''(0)$ with good precision. This means that not only does the PINN ``discover'' by itself the presence of an UV fixed point deformed by a relevant operator, but it also estimates with good precision the number of degrees of freedom at the fixed point, $N_\text{UV}$, and the dimension of the relevant operator, $\Delta$. These two quantities can be extracted from $V(0)$ and $V''(0)$. Based on these, we find that the relative errors in the values reconstructed by the PINN are in the following ranges for  theories with $1 \leq \phi_M \leq 5$:  
\begin{equation}
    10^{-6} \lesssim \frac{\delta N_\text{UV}}{N_\text{UV}}
    \lesssim 10^{-5} \,, \qquad 
    10^{-3} \lesssim \frac{\delta \Delta}{\Delta}
    \lesssim 10^{-2} \,.
\end{equation} 
In the case of $V''(0)$, this error is illustrated in \fig{noBCforV}(middle). \fig{noBCforV}(bottom) compares the reconstructed equations of state when the PINN is or is not provided with the boundary conditions \eqref{Vconditions}. We see that the result is more precise in the first case. This  illustrates the high sensitivity of the $S(T)$ curve to the value of $V''(0)$ or, equivalently, to the dimension of the relevant operator that triggers the RG flow. 

Overall, we regard the results summarized in the previous paragraphs  as a remarkable success given the challenges implied by the multi-entangled and multi-scale nature of the problem on the gravity side. Multi-entanglement refers to the fact that the value of the potential at a certain point, $V(\phi=\phi_0)$, affects the thermodynamic properties of all the black brane geometries with horizons such that \mbox{$\phi_H \geq \phi_0$}. In other words, a single potential must be compatible with an infinite number of solutions with different boundary conditions. This feature makes two of the ingredients that we have implemented indispensable. One is the architecture based on two independent NNs (see \fig{fig:gaussian architecture}), which allowed us to determine both the spacetime geometries and the potential at the same time. The other is the use of solution bundles (see Sec.~\ref{bundle}), which allowed us to feed the PINN the information about all the boundary conditions simultaneously.

\begin{figure}
    \centering
    \includegraphics[width=0.95\textwidth]{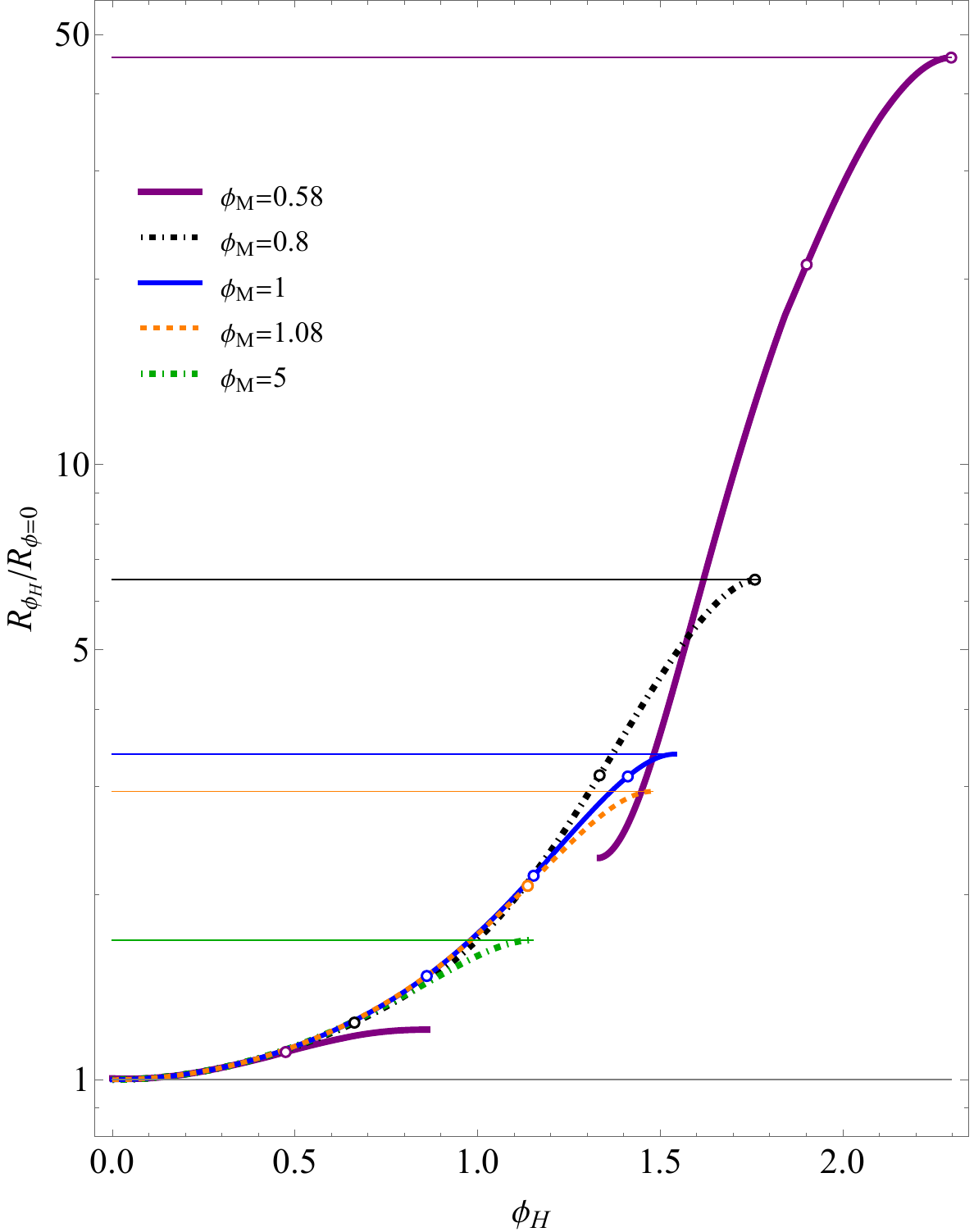}
    \caption{The Ricci scalar at the horizon, normalized to the Ricci scalar at the boundary, varies with the temperature of the solution, i.e.~with $\phi_H$. For theories with a first-order phase transition, the three circles on the curve indicate the three different solutions at the critical temperature. These points coalesce into a single one for a second-order phase transition, and disappear for a crossover. The horizontal lines indicate the endpoints of each curve, which correspond to the zero-temperature, ground-state solutions of each theory. The hierarchy of curvatures increases as we move from a crossover to a second- to a first-order phase transition. The missing intermediate part of the $\phi_M=0.58$ curve is due to the fact that some values of $\phi_H$ do not correspond to thermal states in the dual gauge theory \cite{Bea:2018whf}.}
    \label{curvature}
\end{figure}

The multiscale aspect refers to the fact that the five-dimensional geometries can develop a large hierarchy of scales between the boundary and the horizon. In the gauge theory, this is due to the fact that the RG flow can generate a large hierarchy between the number of degrees of freedom in the UV and in the IR. This aspect is illustrated in \fig{curvature}, where we plot the ratio between the spacetime curvatures at the horizon ($\phi=\phi_H$) and at the boundary ($\phi=0$), as measured by the corresponding Ricci scalars. The intermediate points on some curves indicate solutions with $T=T_c$. The horizontal lines mark the endpoints of each curve, which correspond to the zero-temperature, ground-state solutions. In order to reconstruct correctly the part of the potential that gives rise to the phase transition, the PINN must be able to resolve the hierarchy associated to the intermediate points. In order to reconstruct the deep IR, the PINN must resolve the even larger hierarchy associated to the endpoints of the curves.  
We see that these hierarchies increase as we move from a crossover to a second- to a first-order phase transition. 

\begin{figure}
    \centering
    \includegraphics[width=0.9\textwidth]{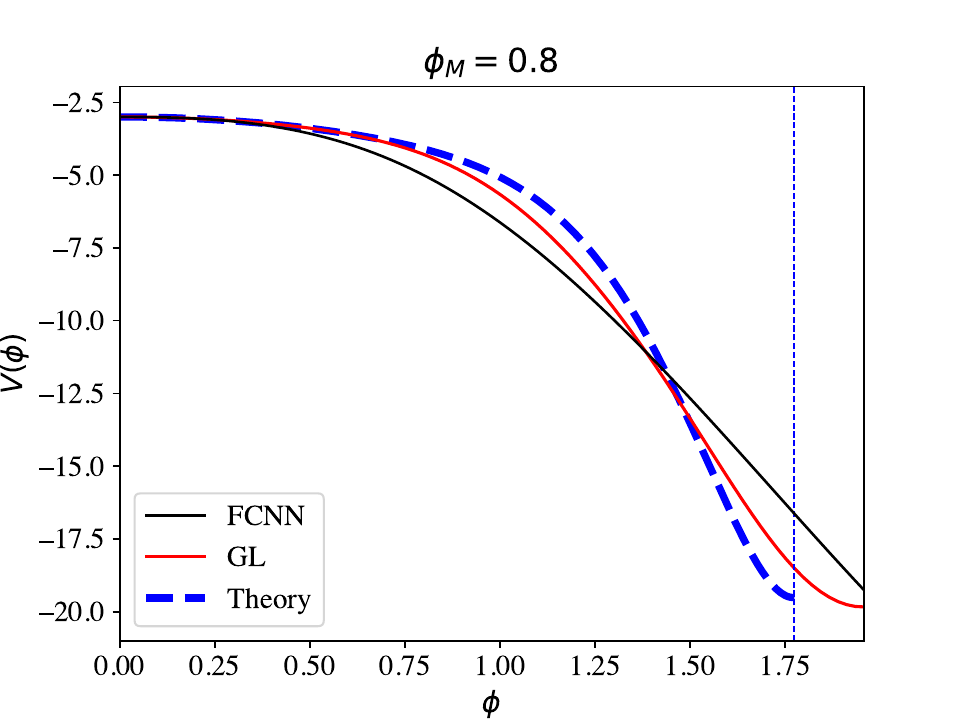} 
    \caption{Comparison between the reconstructed potentials for a theory with $\phi_M=0.8$ with (GL) and without (FCNN) Gaussian localization.}
    \label{GLvsFCNN}
\end{figure}

\begin{figure}
    \centering
    \includegraphics[width=0.95\textwidth]{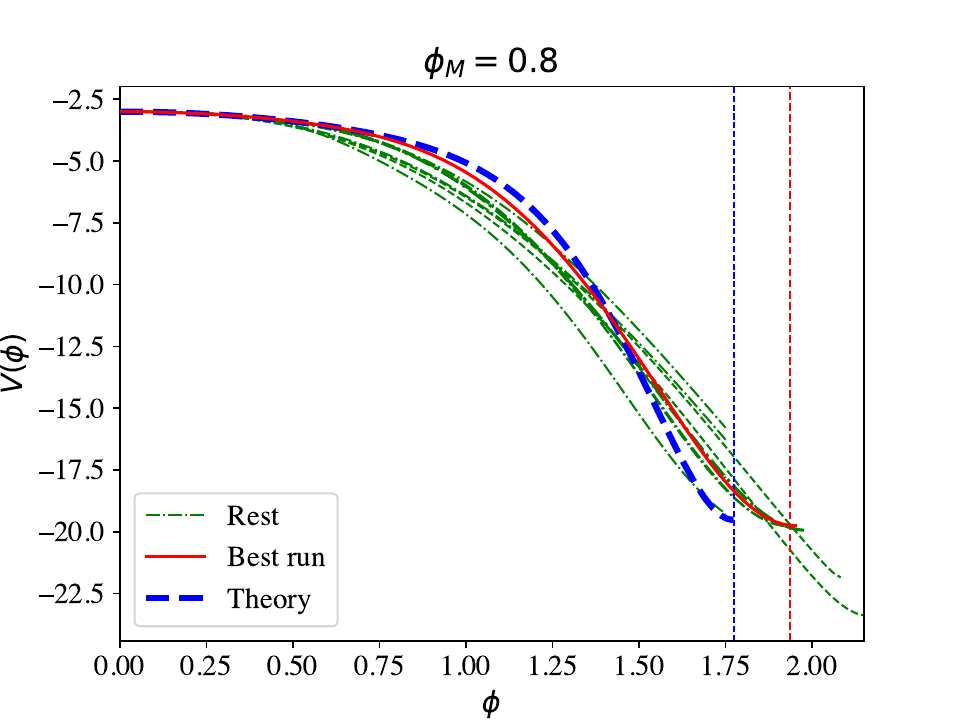} \\
    \includegraphics[width=0.95\textwidth]{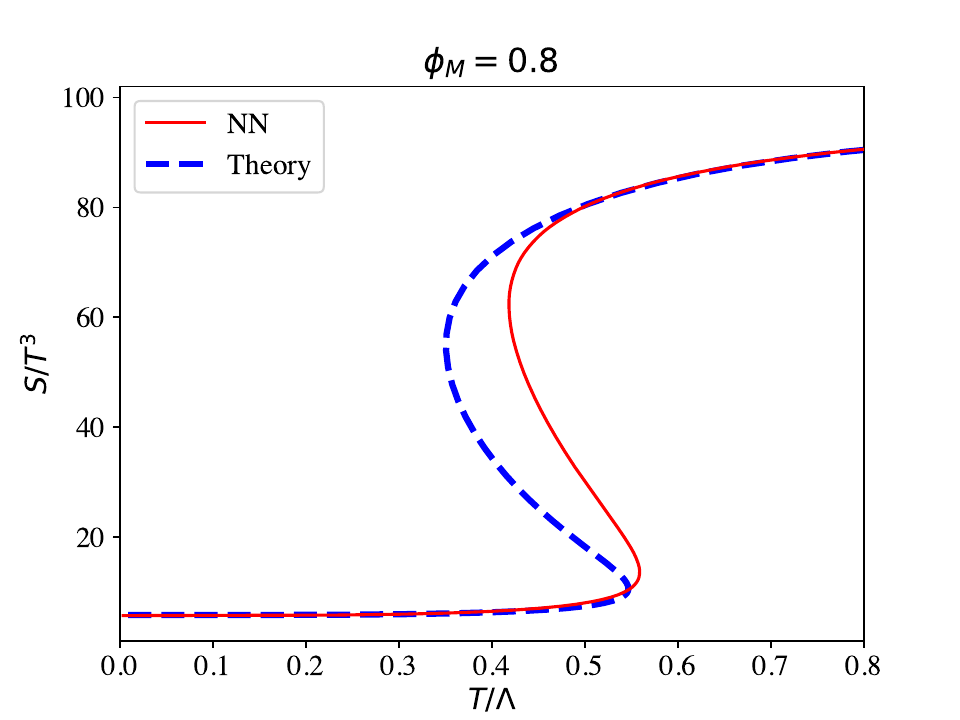}
    \caption{Reconstructed potential (top) and equation of state (bottom) for a theory with a stronger first-order phase transition than that in \fig{Comparison1stOrder}.}
    \label{challenge}
\end{figure}

As with most numerical methods, we expect that the appearance of a separation of scales will make the problem more challenging. The new feature  of ``Gaussian localization''  (GL) that we have developed (see Sec.~\ref{subsubsec: gaussian localization}) helps address this challenge. Intuitively, it results in the specialisation of each neuron on a specific part of the potential. In order to illustrate its effect, in \fig{GLvsFCNN} we compare the  potentials for a theory with $\phi_M=0.8$ that were reconstructed by a NN with GL and by a Fully Connected Neural Network (FCNN) with no GL. The improvement in the results with GL can be seen with the naked eye.  
Nevertheless, theories like the $\phi_M=0.8$ theory, in which the hierarchy of scales is sufficiently large, remain challenging for our method even with GL. This can already be seen by comparing \fig{GLvsFCNN} to \fig{Comparison1stOrder}(top). To illustrate it further, in \fig{challenge}(top)
we show the additional 9 runs that led to the reconstructed potential with GL in \fig{GLvsFCNN}. We see that the dispersion is much larger than in the $\phi_M=1$ case of \fig{Comparison1stOrder}(top). Finally, in \fig{challenge}(bottom) we show the reconstructed  equation of state. Although the presence of a first-order phase transition is recovered, the quality of the result is clearly inferior to that in \fig{Comparison1stOrder}(bottom). Motivated by these challenges, we are currently improving our algorithm by implementing transfer learning. This will allow the PINN to use  an already-determined solution to be used as a seed to find the solution for a harder problem. 

The above discussion provides the right perspective with which to frame our results. On the one hand, we regard them  as an important proof of concept that the gravitational theory can be reconstructed from the equation of state of the dual gauge theory. On the other hand, the previous paragraph illustrates that much work is still necessary in order to perfect the method. Once this is achieved, we expect far-reaching implications for a wide range of physical systems including, but not limited to, those described by gauge theories. Indeed, although we have focused on the thermodynamics of the latter, the equation of state is one of the most fundamental properties of any physical system. Since it only requires knowledge of the equilibrium properties, it can often be determined with a variety of methods. In contrast, the out-of-equilibrium dynamics of many physical systems remains an outstanding challenge. In fact, the exact simulation of this dynamics may be beyond the reach of classical calculations and require quantum computers \cite{Bauer:2022hpo,Bauer:2023qgm}. Given the scale of this challenge,  the gravitational dual provided by holography may provide a valuable, if not the only, analytical approximation to the out-of-equilibrium dynamics of many systems. 

\acknowledgments
We thank Jorge Casalderrey-Solana and Javier Gomez Subils for very useful discussions. This work was supported by the “Center of Excellence Maria de Maeztu 2020-2023” award to the ICCUB (CEX2019-000918-M) funded by MCIN/AEI/10.13039/501100011033. YB, DM and PeT acknowledge support from grants PID2019-105614GB-C22 and 2021-SGR-872. The work of YB is also funded by a Maria Zambrano postdoctoral fellowship from the University of Barcelona. PeT is supported by the project ``Dark Energy and the Origin of the Universe'' (PRE2022-102220), funded by MCIN/AEI/10.13039/501100011033. Funding for the work of RJ was partially provided by the projects PGC2018-098866-B-I00 and FEDER “Una manera de hacer Europa”. The work by PaT is supported by the project ``Accurate Cosmology and the Laws of Nature'' (PID2022-141125NB-I00) with grant PREP2022-000507, funded by MCIN/AEI/10.13039/501100011033.

\appendix
\section{High-temperature behaviour}
\label{appe}
Consider a four-dimensional CFT deformed by a relevant operator $\mathcal{O}$ of conformal dimension $\Delta$ with source $\Lambda$. The action takes the form 
\begin{equation}
    S = S_\text{CFT} + \int d^4x \, 
    \Lambda \mathcal{O} (x) \,.
\end{equation}
For consistency, $\Lambda$ must have dimension $4-\Delta$. 
For homogeneous, thermal equilibrium states, the trace Ward identity takes the form 
\begin{equation}
\mathcal{E} - 3 \mathcal{P} = \Lambda 
\langle \mathcal{O} \rangle \,,
\label{Ward}
\end{equation}
with $\mathcal{E}$ and $\mathcal{P}$ the energy density and the pressure, respectively. The existence of a conformal, UV fixed point implies that the trace of the stress tensor must vanish at leading order at high temperatures. In this limit,  all thermodynamic quantities scale with the temperature as dictated by dimensional analysis, so we have:
\begin{equation}
    \mathcal{E} \simeq 3 \mathcal{P} \sim T^4 \,.
    \label{leading}
\end{equation}
Assuming that $\langle \mathcal{O} \rangle$ vanishes in the undeformed theory, its value at leading order in the deformed theory must be linear in the deformation parameter $\Lambda$, i.e.
\begin{equation}
    \langle \mathcal{O} \rangle \propto \Lambda \, T^{2\Delta-4}\,,
\end{equation}
where the power of $T$ is fixed by dimensional analysis. Through the Ward identity \eqref{Ward}, this fixes the leading-order correction to the asymptotic behaviour \eqref{leading} and, together with the thermodynamic identity $S=\partial \mathcal{P} / \partial T$, it  results in the entropy density
\begin{equation}
    S = c_1 \, T^3 + c_2 \, T^{2\Delta -5} +\cdots \,,
    \label{leadingS}
\end{equation}
where $c_1, c_2$ are numerical coefficients. The main conclusion is that the dimension of the operator responsible for triggering the flow is encoded in the thermodynamic curve $S(T)$. In particular, $\Delta$ can be extracted from the subleading term in the  high-temperature expansion of the equation of state.

\bibliographystyle{JHEP}
%\bibliography{holobib.bib,references}

\providecommand{\href}[2]{#2}\begingroup\raggedright\endgroup

\end{document}